\documentclass[a4paper,11pt]{article}

\usepackage{amsmath,bm}    
\usepackage{amssymb}    
\usepackage{graphicx}   
\usepackage{natbib}    
\usepackage{epsf}
\usepackage{lscape}
\usepackage{subfigure}
\usepackage{enumerate}
\usepackage{color}
\usepackage{algorithm}
\usepackage[]{algorithmic}
\usepackage[utf8]{inputenc}
\usepackage[T1]{fontenc}
\bibpunct{(}{)}{;}{a}{,}{,}
\usepackage[margin=1in]{geometry}

\usepackage{paralist}
\usepackage{xr}

\newcommand{\R}{\mathbb R}

\newtheorem{theorem}{Theorem}
\newtheorem{lemma}[theorem]{Lemma}
\newtheorem{corollary}[theorem]{Corollary}

\newtheorem{example}{Example}
\newtheorem{remark}{Remark}
\providecommand{\keywords}[1]{\textbf{\text{Keywords:}} #1}
\newcommand{\half}{$\left(-{\frac{1}{2}} , {\frac{1}{2}}\right)$ }

\begin{document}

\title{Natural (non-)informative priors for skew-symmetric distributions}

\author{Holger Dette$^a$, Christophe Ley$^b$, and Francisco J. Rubio$^c$
\\ \small $^{a}$ Ruhr-Universit{\"a}t Bochum, Germany.
\\ \small $^{b}$ Ghent University, Belgium.
\\ \small $^{c}$ London School of Hygiene \& Tropical Medicine, UK.}

\date{}  
\maketitle

\begin{abstract}
In this paper, we present an innovative method for constructing proper priors for the skewness (shape) parameter in the skew-symmetric family of distributions. The proposed method is based on assigning a prior distribution on the perturbation effect of the shape parameter, which is quantified in terms of the Total Variation distance. We discuss strategies to translate prior beliefs about the asymmetry of the data into an informative prior distribution of this class.  We show via a Monte Carlo simulation study that our  noninformative priors induce posterior distributions with good frequentist properties,  similar to those of the Jeffreys prior. Our informative priors yield better results than their competitors from the literature. We also propose a scale- and location-invariant prior structure for models with unknown location and scale parameters and provide sufficient conditions for the propriety of the corresponding posterior distribution. Illustrative examples are presented using simulated and real data.
\end{abstract}

\section{Introduction}\label{intro}
It is a well-known fact that several data sets cannot be modeled by means of symmetric distributions, and hence even less via the normal distribution, due to skewness inherent to the data.  Such data are frequently encountered in domains such as  biometry, finance, materials sciences or environmetrics, to cite but these. See for instance \cite{L15} for detailed explanations.

Given these needs, there exists a plethora of distinct proposals for skew distributions in the literature; for a  recent and extensive overview of the state-of-the-art, we refer the reader to the discussion paper \cite{J15}.  A popular class of such distributions are the
\emph{skew-symmetric densities} of the form
\begin{equation}\label{sspdf}
s_{f;G}(x;\mu,\sigma,\lambda) = \frac{2}{\sigma}f\left(\frac{x-\mu}{\sigma}\right)G\left(\lambda \,\omega\left(\frac{x-\mu}{\sigma}\right)\right), \,\,\, x\in {\mathbb R},
\end{equation}
with $f$ the symmetric density (to be skewed), $G$ any symmetric, univariate, absolutely continuous cumulative distribution function (cdf), and $\omega$ an  odd function \citep{AC03,W04}. In \eqref{sspdf}, $\mu\in\R$ is a location, $\sigma\in\R_0^+$ a scale, $\lambda\in\R$ a skewness parameter, and we omit the dependence on $\omega$ in $s_{f;G}$ for the sake of notation. These distributions generalize the popular \emph{skew-normal}  distribution, corresponding to $f$ and $G$ respectively the density and cdf of the standard normal distribution and $\omega$ the identity function, which was introduced in the seminal paper \cite{A85}.  For a recent account on skew-symmetric distributions and, in particular, the skew-normal distribution, we refer the reader to the monograph~\cite{AC14}. We focus on the study of skew-symmetric models of type \eqref{sspdf}  where $\omega$ is positive on $\R^+$ in order to be able to identify right- and left-skewness with the sign of $\lambda$ and to obtain general results. 

Bayesian inference within these families is a  challenge. The prior elicitation for $\lambda$ is complicated since  this parameter not only controls the asymmetry  of~\eqref{sspdf} but also the  mode,  spread, and  tail behaviour.  Numerous priors for $\lambda$ have been proposed in the literature, \emph{inter alia} by  \cite{LL06}, \cite{C12}, \cite{BGL13} and  \cite{RL14}. These references focus on the construction of ``noninformative priors'' from different viewpoints. However there are several situations where we do have \emph{a priori} information on how the data shall behave, and hence at least we know the sign of  $\lambda$. For instance, when modeling BMI (body-mass index) data, we know the data will be right-skewed for biometric reasons, see \textit{e.g.}~\cite{H03}. The same holds true for other biometric indicators and size measurements. Given the popularity of skew-symmetric distributions it is thus of paramount importance to construct informative priors for $\lambda$ that reflect our \emph{a priori} knowledge of the situation. To our knowledge, only \cite{CS13}, who proposed the use of normal and skew-normal priors for $\lambda$ in the skew-normal model, have studied informative priors. Their main motivation for using these kinds of priors is that they facilitate sampling from the corresponding posterior distribution. Their skew-normal prior requires the specification of 3 hyperparameters, which is a difficult task given the interdependence of the parameters. The prior elicitation proposed by \cite{CS13} also relies in the interpretation of $\lambda$ as a parameter regulating solely the asymmetry, hence ignores its overall  effect on the density.

In the present  paper we  tackle the problem of constructing priors for $\lambda$ by interpreting it as a \emph{perturbation parameter}  turning the initial symmetric density $f$  into a skew-symmetric density of the form \eqref{sspdf}. Indeed $f$ is modified by multiplication with a ``skewing function'' $2G(\lambda \,\cdot)$, which is also  referred to as ``modulation of symmetry'' (see,  {e.g.},~\citealp{AC14}). This perturbation effect becomes obvious when we consider $\lambda=0$: only then do we retrieve the initial (symmetric) density $f$, while any non-zero value of $\lambda$ induces a perturbation. Viewing $\lambda$ as perturbation parameter actually reflects its very nature as foreseen by  Fernando de Helguero (1880--1908), the early pioneer of skew-symmetric
distributions. Quoting him ``\emph{But it may happen, and indeed this must often take place, that other perturbation causes join in [...] The curve will be abnormal, asymmetrical''}\footnote{This is a passage from \cite{dH09} translated to English in \cite{AR12}.}.

With this interpretation of $\lambda$ as perturbation parameter it is appealing to invoke its perturbation capacity as a principle on which to construct prior distributions. In Section \ref{sec:distances} we shall therefore measure this effect of $\lambda$ by calculating the Total Variation distance between $f$ and its skew-symmetric counterpart~\eqref{sspdf}. Rather than putting a prior on the parameter $\lambda$, whose values are difficult to interpret, we shall put a prior on this interpretable distance. We opt in Section \ref{ProposedPriorSect} to assign  Beta distributions on the range of values taken by this distance. This allows us, by varying the choice  of the Beta hyperparameters, to build informative as well as noninformative priors, which moreover enjoy a clear interpretability. In Section \ref{ExamplesSect}, we first compare the performance of our priors to existing priors by means of a Monte Carlo simulation study, and then we illustrate their usefulness by analyzing a data set. Finally some proofs  are provided in the Appendix. The present paper is complemented by an online Supplementary Material containing further details on the simulation study and a short application of our methodology to other  distributions containing a shape parameter.

\section{Measuring the perturbation within skew-symmetric families}\label{sec:distances}

There exist several distinct measures for the distance  between two distributions. Those are called probability distances (or metrics, if the distance happens to be a true metric, see \citealp{GS02}). Our choice in the present paper for the Total Variation metric has been driven by the fact that  this distance allows precisely to measure mass relocation when passing from $f$ to $s_{f;G}$ for a given value of the parameter $\lambda$. Moreover, contrary to other distances such as the Hellinger distance or Kullback-Leibler divergence, the
Total Variation distance  seems tailor-made for the problem at hand as it gives rise to simple expressions which is mostly not the case for other distances but is obviously crucial for our goal of building a prior for $\lambda$.

The Total Variation distance between two probability measures $\mu(\cdot)$ and $\nu(\cdot)$ on ${\mathbb R}$ is defined as
\begin{eqnarray*}
d_{TV}(\mu,\nu)=\sup_{A\subset\R}|\mu(A)-\nu(A)|,
\end{eqnarray*}
explaining why this distance represents the largest possible difference between the probability assigned to the same event by two such measures. One easily sees that $0\leq d_{TV}(\mu,\nu) \leq 1$. If the probability measures  admit Radon-Nikodym derivatives $f_1$ and $f_2$, supported on  the interval ${\mathbb R}$, then the definition becomes
\begin{equation*}
d_{TV}(f_1,f_2)=\dfrac{1}{2} \int_{\mathbb R} \vert f_1(x)-f_2(x)\vert dx.
\end{equation*}
Using this expression, the Total Variation distance between the baseline symmetric density $f$ and its skew-symmetric counterpart $s_{f;G}$ from~\eqref{sspdf}, for fixed $\lambda\in\R$,  can be written as
\begin{eqnarray*}\label{dtvss}
d_{TV}(f,s_{f;G}\vert\lambda) = \dfrac{1}{2}\int_{\mathbb R} \vert 2G(\lambda \omega(x)) -1 \vert f(x) dx
.
\end{eqnarray*}
The symmetry of $G$ implies that $d_{TV}(f,s_{f;G}\vert\lambda) = d_{TV}(f,s_{f;G}\vert-\lambda)$, hence this distance is not a one-to-one function of the parameter $\lambda$. This suggests using as measure of perturbation the quantity
\begin{eqnarray}\label{MTV}
M_{TV}(\lambda) = \operatorname{sign}(\lambda) d_{TV}(f,s_{f;G}\vert \lambda),
\end{eqnarray}
which enjoys some appealing properties. First, for $f$ and $G$ fixed, $M_{TV}(0)=0$, which corresponds to the case $s_{f;G}=f$. Since $\lambda\mapsto M_{TV}(\lambda)$ is monotone increasing (see equation \eqref{MTVClosedForm} below), the largest difference is obtained for $\lambda\rightarrow\pm\infty$, when $s_{f;G}$ converges to the positive/negative half-$f$. This largest difference equals $\pm1/2$, hence $M_{TV}(\lambda) \in (-1/2,1/2)$. Given that we only consider the case when $f$ and $s_{f;G}$ have the same location and scale parameters, it follows that this measure is also invariant under affine transformations. By construction, we have that $M_{TV}(\lambda)=-M_{TV}(-\lambda)$. These properties resemble the desirable conditions P.1--P.3 discussed in \cite{AG95} for a measure of skewness. However, the condition P.4 (convexity ordering) in \cite{AG95} is not satisfied in general since we are measuring the perturbation effect of  $\lambda$ only with respect to the symmetric baseline density~$f$. Thus, $M_{TV}$ should not be interpreted as a measure of skewness, but well as a function that quantifies the overall perturbation effect of the  parameter $\lambda$.

By using the symmetry properties of $f$ and $G$, we can re-express \eqref{MTV} as
\begin{equation}\label{MTVClosedForm}
M_{TV}(\lambda) = \frac{1-2S_{f;G}(0;\lambda)}{2},
\end{equation}
where $S_{f;G}$ is the cdf associated with $s_{f;G}$. This expression reveals that, for a fixed choice of $f$ and $G$, $M_{TV}$ is simply a re-scaling of the difference between the mass cumulated on either side of $0$ by the distribution $S_{f;G}$ (since $1-2S_{f;G}(0;\lambda)=\{1-S_{f;G}(0;\lambda)\} - S_{f;G}(0;\lambda)$). Therefore $M_{TV}(\lambda)$ measures the effect of the parameter $\lambda$ in terms of the relocation of mass on either side of the symmetry center of $f$. 

\begin{example}  \label{mvdist}
{\rm
For the skew-normal density we use the standard normal probability density function (pdf) $\phi$  and  cdf $\Phi$ for $f$ and $G$ in \eqref{sspdf}, respectively, and $\omega(x)=x$, and  obtain from \cite{GBR16} and \eqref{MTV} the representation
\begin{equation} \label{MTVSMSN}
M_{TV}(\lambda) = \dfrac{\operatorname{ArcTan}(\lambda)}{\pi},
\end{equation}
for the perturbation measure $M_{TV}$. For the skew-Laplace density (obtained when $f$ and $G$ are the Laplace pdf and cdf, respectively, and $\omega(x)=x$) we have
$$
M_{TV}(\lambda) = \dfrac{1}{2}\dfrac{\lambda}{1+\vert\lambda\vert}.
$$
Finally, let $t_\nu$ and $T_{\nu}$  denote the pdf and cdf of the Student $t$ distribution with $\nu >0 $ degrees of freedom, respectively. The density of the skew-$t$ distribution  with $\nu $ degrees of freedom proposed by \cite{AC03}   is given by
\begin{equation*} \label{tdis}
 \frac{2}{\sigma}t_\nu \Big (\frac{x-\mu}{\sigma}\Big )
 T_{\nu+1}
 \Big (\lambda ( {x-\mu} )  \sqrt{\tfrac{\nu+1}{\nu\sigma ^2 +(x-\mu) ^2}}\Big)
, \,\,\, x\in {\mathbb R}.
\end{equation*}
This distribution is a special case of the class of densities defined in \eqref{sspdf}. In the Appendix, we show that its perturbation measure $M_{TV}$ is also given by \eqref{MTVSMSN} and  therefore coincides with the corresponding  measure for  the skew-normal distribution (which is a special case of the skew-$t$ when $\nu\rightarrow\infty$).
}
\end{example}

\section{Proposed priors}\label{ProposedPriorSect}

The proposed perturbation measure $M_{TV}(\lambda)$ allows us to build informative as well as non-informative priors for the perturbation parameter $\lambda$ in skew-symmetric models. Recall that  $M_{TV}$ varies in \half and is an injective function of $\lambda$.
Consequently any probability distribution on \half  as prior choice for $M_{TV}$ induces a proper prior on $\lambda$. For these distributions
 we choose  the very versatile beta distribution with density
\begin{eqnarray*}\label{betadist}
{1\over B(\alpha,\beta)}
  \left (u+  {1\over 2}  \right)^{\alpha-1} \left( {1\over 2}-u \right)^{\beta-1} ~, ~~u \in \left( -{1\over 2}, {1\over 2}\right ),
\end{eqnarray*}
where $B(\alpha,\beta)$ represents the beta function and  $\alpha,\beta>0$.   We refer to this class of priors as the Beta Total Variation priors $BTV(\alpha,\beta)$ with hyperparameters $\alpha,\beta>0$. Of course, any other distribution with support  \half
   can be employed instead of the beta distribution, however, this choice facilitates some aspects of our study thanks to its flexibility and interpretability.

Our way of proceeding leads to tractable and interpretable priors. In the previous section we have seen that the perturbation effect of $\lambda$ basically consists of a mass relocation \eqref{MTVClosedForm}. This mass relocation affects the shape of the density differently in distinct skew-symmetric models \eqref{sspdf}. Figure \ref{fig:ShapesMTV} shows the density shapes obtained for several values of the percentage of total relocated mass ($M_{TV}(\lambda)$) in the skew-normal, skew-logistic, and skew-Laplace cases. Thus, the interpretation of the perturbation function $M_{TV}(\lambda)$ together with the aid of visualizing the shape of the density for different values of $M_{TV}(\lambda)$ can be used to translate prior believes about the shape of the density into a prior distribution as follows.
\begin{itemize}
\item {\bf Informative priors}. If, \emph{a priori}, we favour right/left asymmetry and hence need informative priors, we choose the hyperparameters $\alpha$ and $\beta$ in such a way that the beta prior assigns mass to the appropriate range of values of $M_{TV}$.

\begin{figure}[h]
\begin{center}
\begin{tabular}{c c c}
\includegraphics[width=5cm, height=4cm]{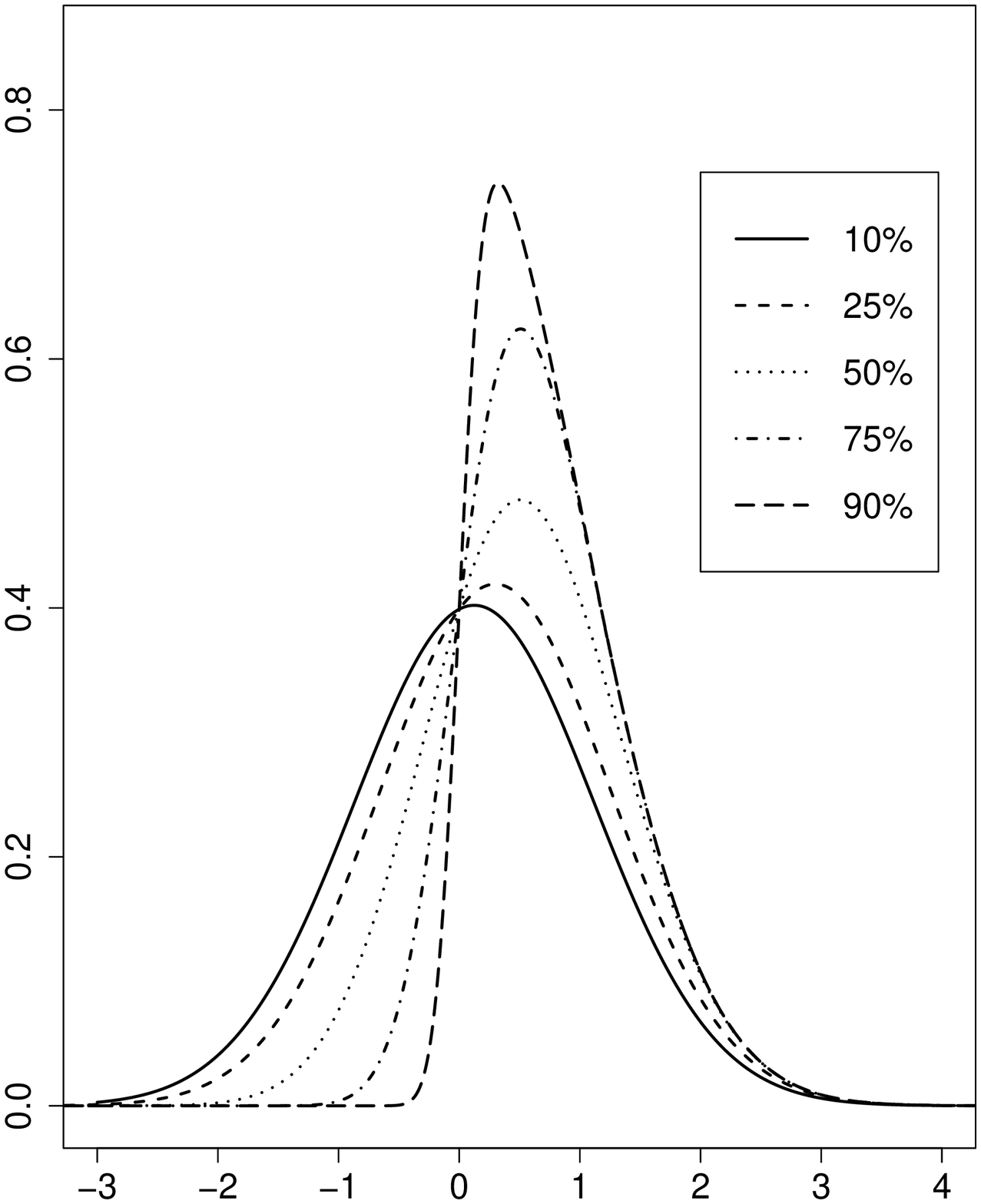}&
\includegraphics[width=5cm, height=4cm]{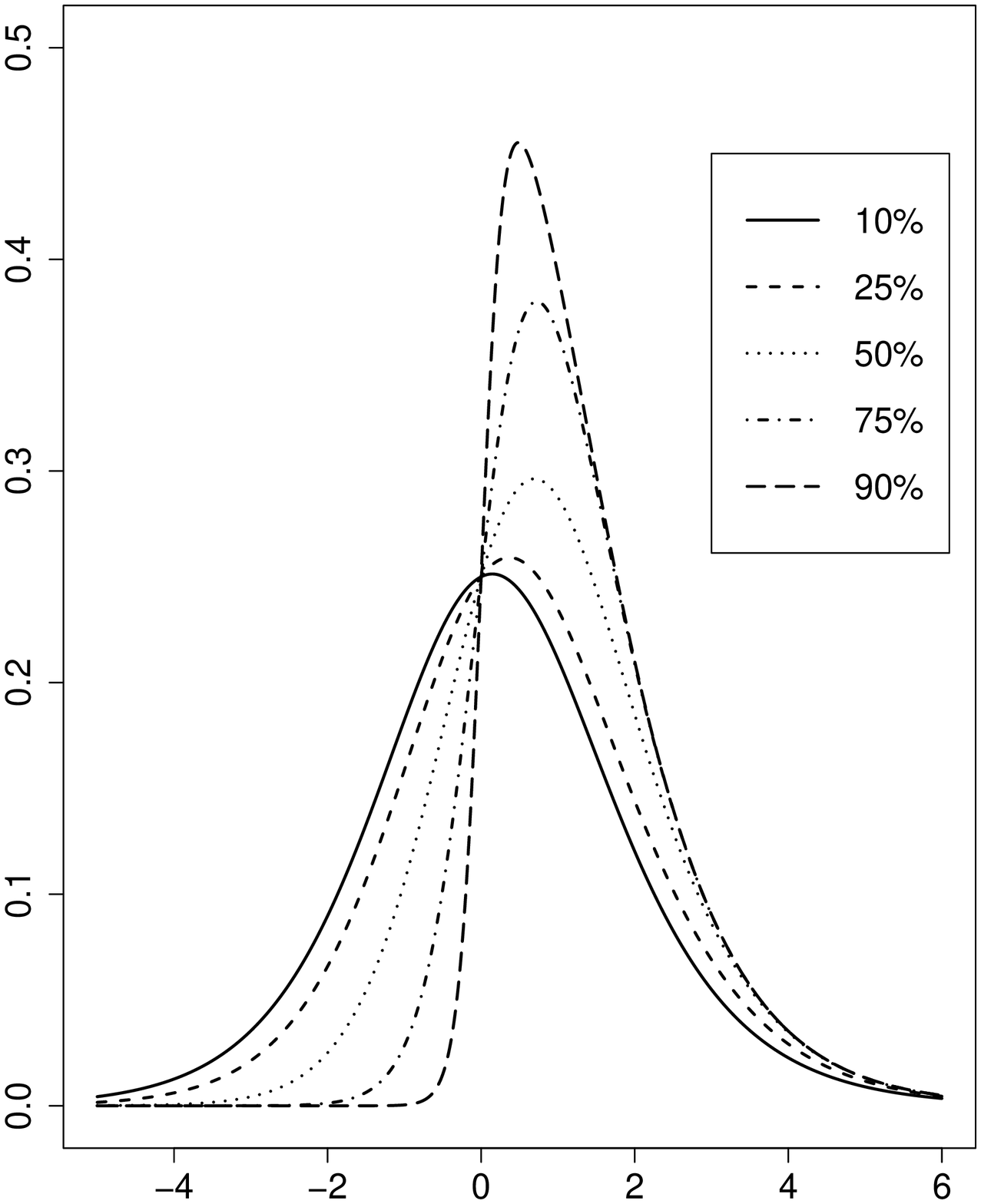}&
\includegraphics[width=5cm, height=4cm]{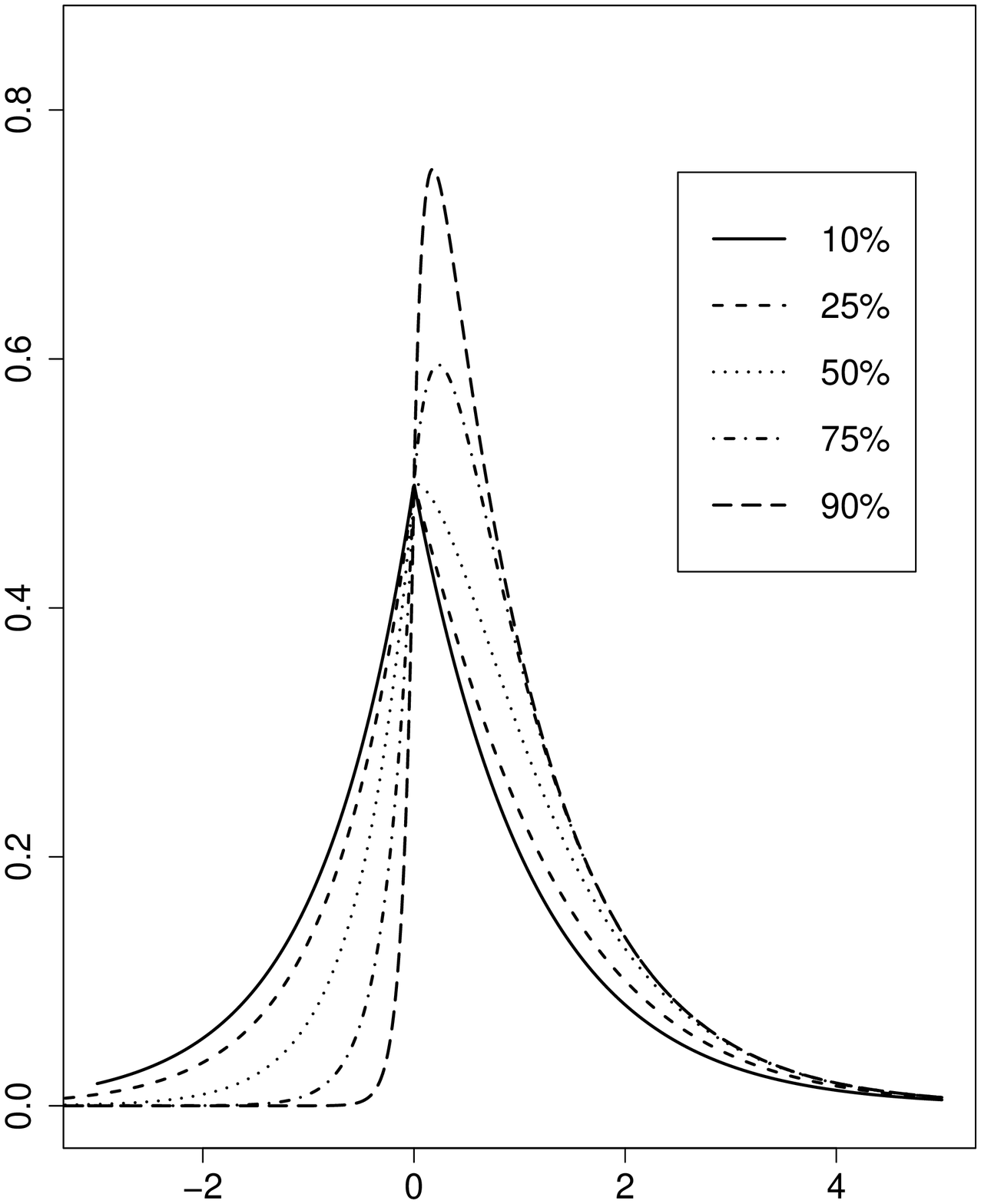}\\
(a) & (b) & (c)
\end{tabular}
\end{center}
\caption{ Shapes of the density for $\mu=0$, $\sigma=1$ and the percentage of total relocated mass equal to $10\%, 25\%, 50\%, 75\%, 90\%$: (a) skew-normal; (b) skew-logistic; and (c) skew-Laplace.}
\label{fig:ShapesMTV}
\end{figure}

\item {\bf Non-informative priors}. For those cases where there is no reliable prior information about the asymmetry of the data, we explore the use of two types of noninformative priors, obtained for (i) $\alpha=\beta=1$, the uniform distribution, which gives equal probability mass to any pair of subintervals of $[0,1]$ of equal length,  and (ii) $\alpha=\beta=1/2$, corresponding to a U-shape beta density. The second choice is motivated as follows. By assigning a $Beta(\alpha,\beta)$ prior to an interpretable measure of perturbation, we implicitly associate a probability $p$ with values that produce right-skewed distributions, and a probability $1-p$ with values that produce left-skewed distributions. We can interpret this scenario as a Bernoulli trial with parameter $p$. A noninformative prior that has been widely studied for the parameter $p$ of the Bernoulli distribution is the Jeffreys prior, which is precisely the $Beta(1/2,1/2)$ prior. This is, the idea is to assign noninformative (or vaguely informative) priors to an interpretable function of the shape parameter $\lambda$. This strategy has been discussed in a more general framework in \cite{SSS12}. 
\end{itemize}
In the remainder of this section, we shall first describe and investigate the resulting $BTV(\alpha,\beta)$ priors for the location-scale-free densities $2f(x)G(\lambda \omega(x))$ (Section~\ref{sec:BTV}), and then discuss joint location-scale-skewness priors for the skew-symmetric models of interest~\eqref{sspdf} (Section~\ref{sec:locsca}). A simple remark on the invariance of these sorts of priors is presented below.
\begin{remark}
{\rm
The $BTV(\alpha,\beta)$ priors are invariant under one-to-one transformations of $\lambda$. This implies that the BTV priors associated to a reparameterization $\alpha= h(\lambda)$, where $h:{\mathbb R}\rightarrow D\subset{\mathbb R}$ is a diffeomorphism, can be derived from the corresponding priors on $\lambda$ using a change of variable.}
\end{remark}

\subsection{Beta-TV priors}\label{sec:BTV}
Putting a $Beta(\alpha,\beta)$ prior on $M_{TV}(\lambda)$ induces a prior on the parameter $\lambda$ with pdf
\begin{equation}\label{THEformula}
\pi_{TV}(\lambda\vert \alpha,\beta)={1\over B(\alpha,\beta)}
  \left (M_{TV}(\lambda)+  {1\over 2}  \right)^{\alpha-1} \left( {1\over 2}-M_{TV}(\lambda) \right)^{\beta-1}\dfrac{d}{d\lambda} M_{TV}(\lambda).
\end{equation}
In order to analyze the general priors $BTV(\alpha,\beta)$, we first investigate some properties of the simpler $BTV(1,1)$  prior  which reduces to $\pi_{TV}(\lambda\vert 1,1) =  \dfrac{d}{d\lambda} M_{TV}(\lambda).$
Sufficient conditions for the well-definiteness of this prior are stated in the following result.
\begin{lemma} \label{lem11}
Consider the class of skew-symmetric densities of the type \eqref{sspdf}. If  $g$ is a bounded pdf and $\int_0^{\infty} \omega(x) f(x)dx <\infty$, the $BTV(1,1)$ prior is well-defined  for all $\lambda$  and given by
\begin{eqnarray}\label{PriorDI}
\pi_{TV}(\lambda\vert 1,1) =  2\int_0^{\infty}  \omega(x)  f(x)g(\lambda  \omega(x)  )dx .
\end{eqnarray}
 \end{lemma}
In the following we provide some general properties of the prior \eqref{PriorDI}, including a characterization of its tails in the important case $\omega(x)=x$.
\begin{theorem}\label{CharactTV}
Consider the class of  skew-symmetric densities of the type \eqref{sspdf}, where $g$ is a bounded pdf and
 $\int_0^{\infty} \omega(x) f(x)dx <\infty$. Then, the prior \eqref{PriorDI} has the following properties:
\begin{enumerate}[(i)]
\item $\pi_{TV}(\lambda\vert 1,1)$ is symmetric about $\lambda=0$.
\item If $g$ is unimodal, then $\pi_{TV}(\lambda\vert 1,1)$  is decreasing in $\vert \lambda \vert$.
\item For $\omega(x)=x$, and under the assumptions that $f$ is unimodal, $f(0)=M<\infty$  and $\int_0^{\infty} x g(x)dx <\infty$, the tails of $\pi_{TV}(\lambda\vert 1,1)$  are of order $O(\vert \lambda \vert^{-2})$.
\end{enumerate}
\end{theorem}
\begin{example}
{\rm Using  expression \eqref{PriorDI} with $\omega(x)=x$ we obtain $
\pi_{TV}(\lambda\vert 1,1) =  1/\left(\pi\left(1+\lambda^2\right)\right)$
as $BTV(1,1)$ prior for  the skew-normal  and  skew-$t$ distributions, and $
\pi_{TV}(\lambda\vert 1,1) =  {1}/({2\left(1+\vert\lambda\vert\right)^2})$
for the skew-Laplace distribution.
}
\end{example}
Thanks to~\eqref{THEformula},  any $BTV(\alpha,\beta)$ prior possesses a nice closed-form expression whenever the $BTV(1,1)$ prior does. The  following result describes the tail behaviour of the density  $\pi_{TV}(\lambda\vert \alpha,\beta)$  of the   $BTV(\alpha,\beta)$ prior  and is a consequence of Theorem \ref{CharactTV} and the tail behaviour of Beta-transformations of symmetric distributions, see  Section 4.5 of \cite{J04}.

\begin{corollary} Consider the skew-symmetric densities defined by \eqref{sspdf} for $\omega(x)=x$, together with the assumptions of Theorem~\ref{CharactTV}(iii).
 The right tail of $\pi_{TV}(\lambda\vert \alpha,\beta)$ is of order $O(\vert \lambda \vert^{-\beta-1})$, while its left tail is of order $O(\vert \lambda \vert^{-\alpha-1})$. Moreover, if $\alpha=\beta$, then $\pi_{TV}(\lambda\vert \alpha,\beta)$ is symmetric.
\end{corollary}
In particular, for the $BTV(1/2,1/2)$ prior we obtain the following expression:
\begin{eqnarray}\label{BetaTV}
\pi_{TV}(\lambda\vert 1/2,1/2) = \dfrac{1}{\pi\sqrt{\frac14 -M_{TV}^2 (\lambda)}}\pi_{TV}(\lambda\vert 1,1).
\end{eqnarray}
This prior is symmetric and, for skew-symmetric models with $\omega(x)=x$, its tails are of order $O(\vert \lambda \vert^{-3/2})$, which interestingly coincide with those of the Jeffreys prior \citep{RL14}. However, the prior $\pi_{TV}(\lambda\vert 1/2,1/2)$ and the Jeffreys prior are not identical. In fact, the Jeffreys prior has no closed-form expression, and moreover it can be ill-defined for certain combinations of $f$ and $G$ due to singularities in the Fisher information matrix in the neighborhood of $\lambda=0$, see \cite{HL12}.

\subsection{Heuristic approximations to the $BTV(1,1)$ priors}\label{benadering}
\begin{figure}[h]
\begin{center}
\begin{tabular}{c c}
\includegraphics[width=4cm, height=3cm]{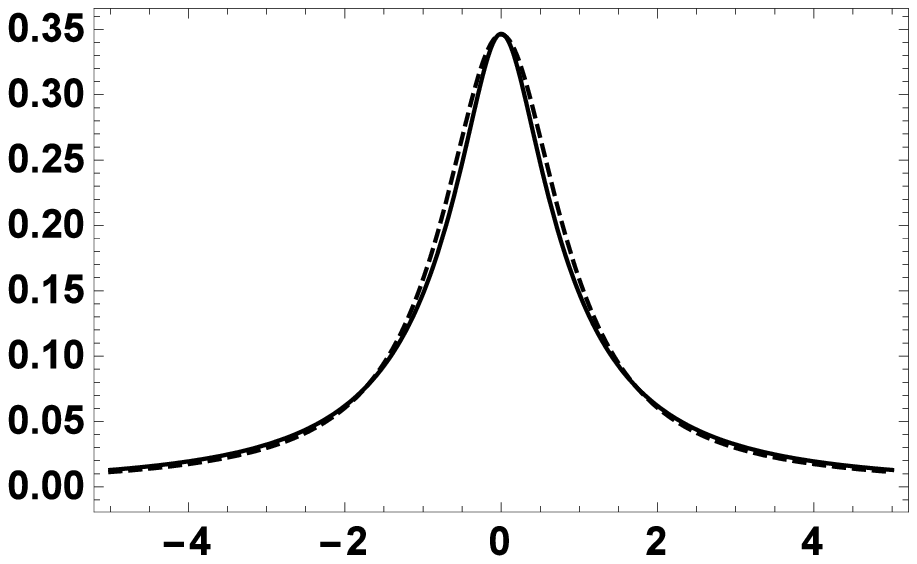}&
\includegraphics[width=4cm, height=3cm]{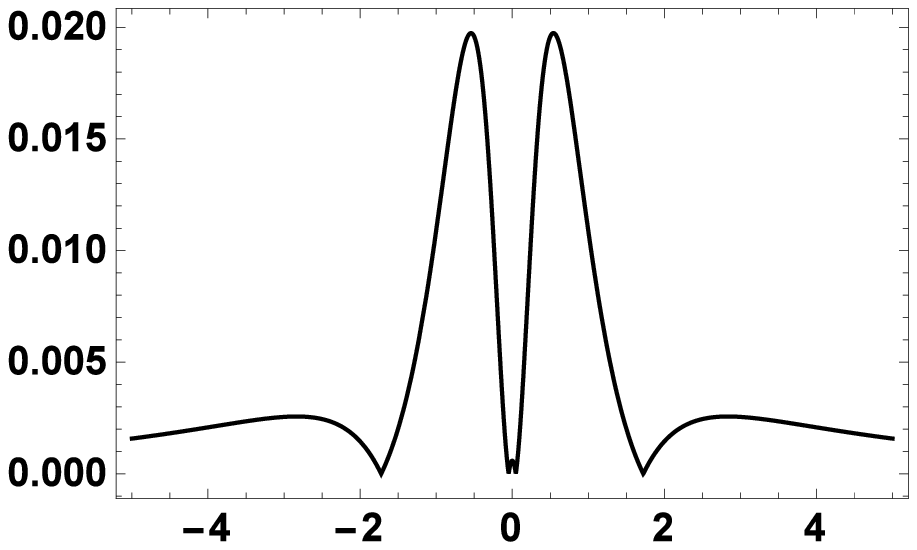}\\
(a) & (b) \\
\end{tabular}
\end{center}
\caption{ (a) Total variation prior of $\lambda$ (continuous line) and Student-$t$ approximation (dashed line); (b) Absolute difference between the Total variation prior of $\lambda$ and the Student-$t$ approximation.
}
\label{fig:approx}
\end{figure}

In general the  expression \eqref{PriorDI} is not available in closed-form. However, we can appeal to the characterization of the tail behaviour of these priors in Theorem \ref{CharactTV} to come up with tractable approximations. For example, in the case when $\omega (x)=x$ and
$f$ and $G$ are the logistic pdf and cdf, respectively,  the $BTV(1,1)$ prior  is not  available in closed-form
but  can be reasonably well approximated with a Student-$t$ distribution with $1$ degree of freedom and scale parameter $0.92$. Figure \ref{fig:approx} shows the quality of this approximation. The quality of Student-$t$ approximations for $BTV(1,1)$ priors associated to other skew-symmetric models seems to require a case by case analysis.

\subsection{Location-scale-skewness models: partial information priors}\label{sec:locsca}

Consider now the initial densities of interest~\eqref{sspdf}, which contain unknown location and scale parameters. For this model we adopt the prior structure
\begin{eqnarray}\label{PriorLC}
\pi(\mu,\sigma,\lambda) = \dfrac{p(\lambda)}{\sigma},
\end{eqnarray}
where $p(\lambda)$ is a proper prior on $\lambda$, here $BTV(\alpha,\beta)$.  This prior structure can be justified as a sort of \emph{partial information prior} \citep{SB98} in the sense that we are using the reference prior for the location and scale parameters, $\pi(\mu,\sigma)\propto \sigma^{-1}$, while we allow for using a subjective prior on the perturbation parameter $\lambda$. Such structures can also be motivated as priors inspired by the form of the independence Jeffreys prior \citep{RS14,RS15}.  Theorem~\ref{PointObs} below presents sufficient conditions for the propriety of the posterior distribution under the prior structure \eqref{PriorLC}. We restrict our study to the cases when $f$ belongs to  scale mixtures of normals. This is a wide family of symmetric distributions which contains many models of practical interest such as the normal, logistic, Laplace, symmetric hyperbolic, Student-$t$, among many other distributions.
\begin{theorem}\label{PointObs}
Let ${\bf x}=(x_1,\dots,x_n)$ be an {i.i.d.} sample from a skew-symmetric model~\eqref{sspdf}. Suppose that $f$ is a scale mixture of normals. Then the posterior distribution of $(\mu,\sigma,\lambda)$ associated with the prior structure (\ref{PriorLC}) is proper if $n\geq 2$ and if all the observations are different.
\end{theorem}
This theorem, proved in the Appendix, guarantees that the priors proposed in the present paper for skew-symmetric densities lead almost surely to proper posterior distributions.

\section{Finite sample properties and practical performance}\label{ExamplesSect}

\subsection{Monte Carlo simulation study}
\subsection*{Noninformative priors}
In  order to compare the performance of the priors proposed in Section \ref{ProposedPriorSect}  with that of the Jeffreys prior (\citealp{LL06}, \citealp{RL14}), we have conducted a thorough simulation study, of which we only present certain results here, the others being provided in the Supplementary Material. We have generated  $N=1,000$ samples of size $n=50$ from the skew-normal,  skew-logistic and skew-Laplace distributions with location parameter $\mu=0$, scale parameter $\sigma=1$, and perturbation parameter $\lambda=0,2.5,5$. Results for the sample sizes $n=100$ and $n=200$ can be found in the Supplementary Material. For each of these samples, we simulate a posterior sample of size $1,000$ from $(\mu,\sigma,\lambda)$ using the  $BTV(1,1)$, $BTV(1/2,1/2)$ and Jeffreys priors. We employ a self-adaptive MCMC sampler \citep{CF06} to obtain the posterior samples. For each posterior sample, we calculate the coverage proportions of the 95\% credible intervals of each parameter (that is, the proportion of credible intervals that contain the true value of the parameter) as well as the 5\%, 50\% and 95\% quantiles of the posterior medians and maximum \emph{a posteriori} (MAP) estimators. In addition, we obtain the median of the Bayes factors (BFs) associated to the hypothesis $H_0:\lambda=0$. The Bayes factors are approximated using the Savage-Dickey density ratio.

The BTV  priors for the skew-normal and skew-Laplace models enjoy nice closed-form expressions.   For the skew-logistic model,  we employ the Student-$t$ approximation for the $BTV(1,1)$ prior described in Section~\ref{benadering}, the $BTV(1/2,1/2)$ prior following then immediately from \eqref{BetaTV}. In order to implement the respective Jeffreys priors, we need to work with approximations. In the skew-normal and skew-logistic cases, we respectively use  the  Student-$t$ approximation proposed in \cite{BB07} (1/2 degrees of freedom and scale $\pi/2$) and the  Student-$t$ approximation proposed in \cite{RL14} (1/2 degrees of freedom and scale 4/3). For the skew-Laplace model, we propose a new approximation to the Jeffreys prior:
\begin{eqnarray*}
\pi_J(\lambda)= \dfrac{1}{4s_0(1+\vert \lambda/s_0\vert)^{3/2}},
\end{eqnarray*}
where $s_0=0.77$.

The results  are reported in Tables \ref{table:SN50}, \ref{table:SLO50} and \ref{table:SL50}.  Overall, we observe that the  $BTV(1/2,1/2)$ and Jeffreys priors exhibit the best, and very similar, performance. This is an expected result as the Jeffreys prior is first-order or second-order probability matching under mild regularity conditions \citep{G11}. However, we emphasize that the $BTV(1/2,1/2)$ prior is more tractable than the Jeffreys prior and it is well-defined
 under less restrictive conditions. These conclusions are further supported by the simulation studies of the Supplementary Material. 

\begin{table}[!htbp]
\begin{center}
\scriptsize
\begin{tabular}[h]{|c| c c c| c c c| c| c|}
\hline
Prior &  \multicolumn{3}{c}{MAP} & \multicolumn{3}{c}{Median} & Coverage & BF \\
\hline
 & 5\% & 50\% & 95\% & 5\% & 50\% & 95\% & &\\
 \hline
$\lambda=0$&&&&&&&&\\
\hline
BTV(1/2,1/2) &&&&&&&&\\
$\mu$ & -1.157 & -0.021 & 1.169 & -0.820 & -0.005 & 0.922 & 0.990 & --\\
  $\sigma$ & 0.904 & 1.103 & 1.493 & 0.992 & 1.195 & 1.515 & 0.874 & --\\
  $\lambda$ & -1.597 & 0.011 & 1.453 & -1.769 & 0.004 & 1.484 & 0.990 & 1.715\\
Jeffreys &&&&&&&&\\
$\mu$ & -1.170 & -0.076 & 1.229 & -0.871 & -0.018 & 1.008 & 0.983 & --\\
  $\sigma$ & 0.923 & 1.115 & 1.541 & 0.999 & 1.216 & 1.528 & 0.858 & --\\
  $\lambda$ & -1.854 & 0.015 & 1.663 & -1.876 & 0.017 & 1.589 & 0.986 & 1.824\\
BTV(1,1) &&&&&&&&\\
$\mu$ & -1.059 & 0.004 & 1.089 & -0.647 & -0.007 & 0.731 & 0.997 & --\\
  $\sigma$ & 0.897 & 1.081 & 1.344 & 0.974 & 1.163 & 1.412 & 0.892 & --\\
  $\lambda$ & -0.712 & 0.003 & 0.552 & -1.158 & -0.011 & 0.938 & 0.996 & 1.245\\
\hline
$\lambda=2.5$&&&&&&&&\\
\hline
BTV(1/2,1/2) &&&&&&&&\\
$\mu$ & -0.281 & 0.039 & 0.921 & -0.224 & 0.189 & 0.821 & 0.899 & --\\
  $\sigma$ & 0.610 & 0.832 & 1.220 & 0.667 & 0.880 & 1.202 & 0.931 & --\\
  $\lambda$ & -0.273 & 1.033 & 5.290 & -0.103 & 1.414 & 7.759 & 0.869 & 0.949\\
Jeffreys &&&&&&&&\\
$\mu$ & -0.283 & 0.036 & 0.994 & -0.233 & 0.170 & 0.837 & 0.897 & --\\
  $\sigma$ & 0.614 & 0.847 & 1.220 & 0.674 & 0.891 & 1.213 & 0.936 & --\\
  $\lambda$ & -0.307 & 1.342 & 5.964 & -0.119 & 1.560 & 8.571 & 0.877 & 0.988\\
BTV(1,1) &&&&&&&&\\
$\mu$ & -0.225 & 0.093 & 0.891 & -0.163 & 0.308 & 0.815 & 0.862 & --\\
  $\sigma$ & 0.602 & 0.782 & 1.171 & 0.647 & 0.845 & 1.147 & 0.917 & --\\
  $\lambda$ & -0.163 & 0.415 & 4.094 & -0.076 & 1.032 & 5.345 & 0.843 & 0.797\\
\hline
$\lambda=5$&&&&&&&&\\
\hline
BTV(1/2,1/2) &&&&&&&&\\
$\mu$ & -0.174 & -0.004 & 0.341 & -0.157 & 0.026 & 0.576 & 0.918 & --\\
  $\sigma$ & 0.594 & 0.958 & 1.197 & 0.662 & 0.960 & 1.193 & 0.926 & --\\
  $\lambda$ & -15.792 & 3.132 & 30.601 & 0.557 & 4.759 & 31.230 & 0.891 & 0.140\\
Jeffreys &&&&&&&&\\
$\mu$ & -0.180 & -0.007 & 0.318 & -0.153 & 0.019 & 0.552 & 0.919 & --\\
  $\sigma$ & 0.595 & 0.958 & 1.200 & 0.666 & 0.963 & 1.199 & 0.925 & --\\
  $\lambda$ & -7.616 & 3.265 & 38.095 & 0.609 & 4.849 & 32.032 & 0.896 & 0.136\\
BTV(1,1) &&&&&&&&\\
$\mu$ & -0.141 & 0.028 & 0.623 & -0.114 & 0.072 & 0.642 & 0.895 & --\\
  $\sigma$ & 0.581 & 0.918 & 1.153 & 0.639 & 0.921 & 1.152 & 0.909 & --\\
  $\lambda$ & -0.010 & 2.921 & 8.071 & 0.344 & 3.595 & 11.755 & 0.874 & 0.164\\
\hline
\end{tabular}
\caption{Skew-normal data for  noninformative priors: $\mu=0,\sigma=1, n=50$.}
\label{table:SN50}
\end{center}
\end{table}

\begin{table}[!htbp]
\begin{center}
\scriptsize
\begin{tabular}[h]{|c| c c c| c c c| c| c|}
\hline
Prior &  \multicolumn{3}{c}{MAP} & \multicolumn{3}{c}{Median} & Coverage & BF \\
\hline
 & 5\% & 50\% & 95\% & 5\% & 50\% & 95\% & &\\
  \hline
$\lambda=0$&&&&&&&&\\
\hline
BTV(1/2,1/2) &&&&&&&&\\
$\mu$ & -1.513 & -0.006 & 1.624 & -1.260 & 0.042 & 1.368 & 0.964 & --\\
  $\sigma$ & 0.871 & 1.080 & 1.350 & 0.922 & 1.148 & 1.432 & 0.905 & --\\
  $\lambda$ & -0.771 & 0.005 & 0.753 & -1.278 & -0.027 & 1.099 & 0.972 & 2.049\\
Jeffreys &&&&&&&&\\
$\mu$ & -1.524 & 0.017 & 1.580 & -1.279 & 0.046 & 1.384 & 0.966 & --\\
  $\sigma$ & 0.871 & 1.089 & 1.372 & 0.927 & 1.153 & 1.433 & 0.901 & --\\
  $\lambda$ & -0.862 & -0.002 & 0.802 & -1.327 & -0.026 & 1.107 & 0.967 & 2.138\\
BTV(1,1) &&&&&&&&\\
$\mu$ & -1.378 & 0.030 & 1.397 & -1.163 & 0.031 & 1.187 & 0.980 & --\\
  $\sigma$ & 0.857 & 1.065 & 1.307 & 0.913 & 1.126 & 1.379 & 0.923 & --\\
  $\lambda$ & -0.675 & 0.002 & 0.629 & -1.009 & -0.022 & 0.920 & 0.983 & 1.436\\
\hline
$\lambda=2.5$&&&&&&&&\\
\hline
BTV(1/2,1/2) &&&&&&&&\\
$\mu$ & -0.459 & 0.080 & 0.810 & -0.390 & 0.175 & 0.924 & 0.913 & --\\
  $\sigma$ & 0.617 & 0.871 & 1.252 & 0.650 & 0.909 & 1.266 & 0.923 & --\\
  $\lambda$ & 0.022 & 1.172 & 7.476 & 0.314 & 1.809 & 11.010 & 0.912 & 0.507\\
Jeffreys &&&&&&&&\\
$\mu$ & -0.476 & 0.065 & 0.815 & -0.378 & 0.162 & 0.935 & 0.912 & --\\
  $\sigma$ & 0.624 & 0.875 & 1.269 & 0.658 & 0.917 & 1.266 & 0.919 & --\\
  $\lambda$ & 0.073 & 1.321 & 7.835 & 0.316 & 1.894 & 9.999 & 0.905 & 0.518\\
BTV(1,1) &&&&&&&&\\
$\mu$ & -0.357 & 0.167 & 0.859 & -0.286 & 0.276 & 0.989 & 0.899 & --\\
  $\sigma$ & 0.606 & 0.835 & 1.204 & 0.641 & 0.875 & 1.196 & 0.899 & --\\
  $\lambda$ & 0.209 & 0.937 & 4.343 & 0.252 & 1.439 & 5.855 & 0.879 & 0.463\\
\hline
$\lambda=5$&&&&&&&&\\
\hline
BTV(1/2,1/2) &&&&&&&&\\
$\mu$ & -0.316 & 0.011 & 0.524 & -0.263 & 0.049 & 0.621 & 0.921 & --\\
  $\sigma$ & 0.614 & 0.938 & 1.230 & 0.656 & 0.955 & 1.236 & 0.919 & --\\
  $\lambda$ & -7.980 & 3.013 & 33.228 & 0.995 & 4.600 & 33.555 & 0.915 &  0.111\\
Jeffreys &&&&&&&&\\
$\mu$ & -0.307 & 0.010 & 0.486 & -0.262 & 0.052 & 0.600 & 0.921 & --\\
  $\sigma$ & 0.618 & 0.938 & 1.230 & 0.662 & 0.958 & 1.242 & 0.919 & --\\
  $\lambda$ & -10.231 & 3.015 & 33.053 & 1.099 & 4.535 & 33.336 & 0.906 & 0.115\\
BTV(1,1) &&&&&&&&\\
$\mu$ & -0.238 & 0.060 & 0.632 & -0.204 & 0.124 & 0.729 & 0.897 & --\\
  $\sigma$ & 0.594 & 0.894 & 1.181 & 0.640 & 0.913 & 1.185 & 0.902 & --\\
  $\lambda$ & 0.543 & 2.585 & 8.449 & 0.826 & 3.483 & 13.135 & 0.894 & 0.109\\
\hline
\end{tabular}
\caption{ Skew-logistic data for  noninformative priors: $\mu=0,\sigma=1, n=50$.}
\label{table:SLO50}
\end{center}
\end{table}

\begin{table}[!htbp]
\begin{center}
\scriptsize
\begin{tabular}[h]{|c| c c c| c c c| c| c|}
\hline
Prior &  \multicolumn{3}{c}{MAP} & \multicolumn{3}{c}{Median} & Coverage & BF \\
\hline
 & 5\% & 50\% & 95\% & 5\% & 50\% & 95\% & &\\
  \hline
$\lambda=0$&&&&&&&&\\
\hline
BTV(1/2,1/2) &&&&&&&&\\
$\mu$ & -0.585 & 0.000 & 0.587 & -0.524 & -0.005 & 0.562 & 0.946 & -- \\
  $\sigma$ & 0.803 & 1.032 & 1.293 & 0.835 & 1.067 & 1.332 & 0.960  & --\\
  $\lambda$ & -0.369 & 0.001 & 0.385 & -0.621 & 0.001 & 0.582 & 0.946  &  3.921\\
Jeffreys &&&&&&&&\\
$\mu$ & -0.591 & -0.007 & 0.629 & -0.523 & -0.009 & 0.575 & 0.948 & --\\
  $\sigma$ & 0.795 & 1.034 & 1.290 & 0.835 & 1.072 & 1.333 & 0.959 & --\\
  $\lambda$ & -0.415 & -0.001 & 0.388 & -0.654 & -0.000 & 0.574 & 0.948 & 3.759\\
BTV(1,1) &&&&&&&&\\
$\mu$ & -0.549 & -0.003 & 0.554 & -0.509 & -0.011 & 0.545 & 0.957 & --\\
  $\sigma$ & 0.798 & 1.031 & 1.286 & 0.824 & 1.065 & 1.318 & 0.958 & --\\
  $\lambda$ & -0.347 & -0.001 & 0.329 & -0.546 & -0.002 & 0.529 & 0.954 &  2.567\\
\hline
$\lambda=2.5$&&&&&&&&\\
\hline
BTV(1/2,1/2) &&&&&&&&\\
$\mu$ & -0.273 & 0.014 & 0.381 & -0.230 & 0.035 & 0.375 & 0.932 & --\\
  $\sigma$ & 0.661 & 0.939 & 1.267 & 0.706 & 0.972 & 1.299 & 0.936 & --\\
  $\lambda$ & 0.280 & 1.558 & 6.824 & 0.697 & 2.234 & 10.721 & 0.926 & 0.137\\
Jeffreys &&&&&&&&\\
$\mu$ & -0.272 & 0.016 & 0.416 & -0.239 & 0.038 & 0.385 & 0.930 & --\\
  $\sigma$ & 0.668 & 0.933 & 1.266 & 0.710 & 0.967 & 1.296 & 0.932 & --\\
  $\lambda$ & 0.270 & 1.493 & 6.402 & 0.684 & 2.194 & 11.039 & 0.922 & 0.136\\
BTV(1,1) &&&&&&&&\\
$\mu$ & -0.229 & 0.041 & 0.421 & -0.192 & 0.063 & 0.402 & 0.931 & --\\
  $\sigma$ & 0.656 & 0.917 & 1.230 & 0.694 & 0.949 & 1.254 & 0.933 & --\\
  $\lambda$ & 0.338 & 1.438 & 4.853 & 0.614 & 1.963 & 7.188 & 0.936 & 0.106\\
\hline
$\lambda=5$&&&&&&&&\\
\hline
BTV(1/2,1/2) &&&&&&&&\\
$\mu$ & -0.187 & 0.000 & 0.244 & -0.164 & 0.015 & 0.262 & 0.931 & --\\
  $\sigma$ & 0.667 & 0.942 & 1.234 & 0.697 & 0.967 & 1.263 & 0.932 & --\\
  $\lambda$ & -3.128 & 3.112 & 30.605 & 1.445 & 4.790 & 34.839 & 0.922 & 0.054\\
Jeffreys &&&&&&&&\\
$\mu$ & -0.182 & 0.002 & 0.257 & -0.166 & 0.019 & 0.264 & 0.929 & --\\
  $\sigma$ & 0.668 & 0.942 & 1.246 & 0.701 & 0.971 & 1.267 & 0.933 & --\\
  $\lambda$ & -6.021 & 3.107 & 30.692 & 1.424 & 4.824 & 34.424 & 0.926 & 0.053\\
BTV(1,1) &&&&&&&&\\
$\mu$ & -0.157 & 0.021 & 0.291 & -0.124 & 0.041 & 0.296 & 0.935 & --\\
  $\sigma$ & 0.658 & 0.914 & 1.211 & 0.688 & 0.944 & 1.237 & 0.933 & --\\
  $\lambda$ & 0.775 & 2.897 & 9.682 & 1.281 & 4.018 & 14.715 & 0.927 & 0.041\\
\hline
\end{tabular}
\caption{Skew-Laplace data for noninformative priors: $\mu=0,\sigma=1, n=50$.}
\label{table:SL50}
\end{center}
\end{table}

\subsection*{Informative priors}
We now explore the use of the proposed informative priors. We simulate $N=1,000$ samples of size $n=50$ from a skew-normal distribution with  parameters $\mu=0$,  $\sigma=1$ and  $\lambda=5$. We employ again a self-adaptive MCMC sampler to obtain the posterior samples. For each of these samples, we simulate a posterior sample of size $1,000$ from $(\mu,\sigma,\lambda)$ using the $BTV(3,1/2)$ prior. This prior assigns  5\% of the mass to values of $\lambda<0$ while being vaguely informative about $\lambda>0$. We also consider the skew-normal prior proposed in \cite{CS13} with hyperparameters $(\mu_0,\sigma_0,\lambda_0)=(0,1,6.5)$. This prior also assigns 5\% of the mass to values of $\lambda<0$ and is vaguely informative about $\lambda>0$, however, it has lighter tails than the BTV  prior. We calculate the coverage proportions of the 95\% credible intervals of each parameter as well as the 5\%, 50\% and 95\% quantiles of the posterior medians and MAP estimators. Results are reported in Table \ref{table:SN50INF}. We observe that the $BTV(3,1/2)$  prior exhibits better frequentist properties than its competitor. This, together with the intuitive nature of our priors, underlines the strength of our new approach.

\begin{table}[!htbp]
\begin{center}
\scriptsize
\begin{tabular}[h]{|c| c c c| c c c| c| }
\hline
Prior &  \multicolumn{3}{c}{MAP} & \multicolumn{3}{c}{Median} & Coverage \\
\hline
 & 5\% & 50\% & 95\% & 5\% & 50\% & 95\% & \\
  \hline
$\lambda=5$&&&&&&&\\
\hline
BTV(3,1/2) &&&&&&&\\
$\mu$ & -0.185 & -0.020 & 0.241 & -0.161 & 0.001 & 0.309 & 0.935 \\
  $\sigma$ & 0.680 & 0.976 & 1.200 & 0.722 & 0.986 & 1.210 & 0.945 \\
  $\lambda$ & -22.182 & 3.348 & 40.942 & 1.546 & 5.269 & 36.241 & 0.919 \\
SN(0,2.5,6.5) &&&&&&&\\
$\mu$ & -0.054 & 0.078 & 0.247 & -0.045 & 0.093 & 0.311 & 0.890 \\
  $\sigma$ & 0.692 & 0.902 & 1.087 & 0.703 & 0.904 & 1.084 & 0.908 \\
  $\lambda$ & 1.118 & 2.962 & 4.147 & 1.494 & 3.091 & 4.204 & 0.804 \\
\hline
\end{tabular}
\caption{Skew-normal data for informative priors: $\mu=0,\sigma=1,n=50$.}
\label{table:SN50INF}
\end{center}
\end{table}

\subsection{Body Mass Index}\label{BMI}

The goal of this section is to investigate the Body Mass Index (BMI) of physically active adult men on basis of data provided in \cite{H03}. We analyse three groups from this data set corresponding to individuals with ages in the ranges: $18-24$, $25-34$, and $35-44$, as these represent parts of the population with similar biological and sociological characteristics. The sample sizes are $n=75,94$ and 45, respectively. Biometric reasons entail that BMI data are typically asymmetric with a longer right tail, and we have consulted an expert on BMI data for our prior specification (Helena Carreira, LSHTM). Consequently, we fit a skew-normal distribution to these data sets together with the prior (\ref{PriorLC}). In cases with small or moderate sample sizes, the profile likelihood of $\lambda$ tends to be flat (see \citealp{AC14}). Thus, we expect informative priors to yield better results since the prior distribution has a relevant effect on the shape of the posterior distribution.

\begin{figure}[h]
\begin{center}
\begin{tabular}{c c c}
\includegraphics[width=5cm, height=4cm]{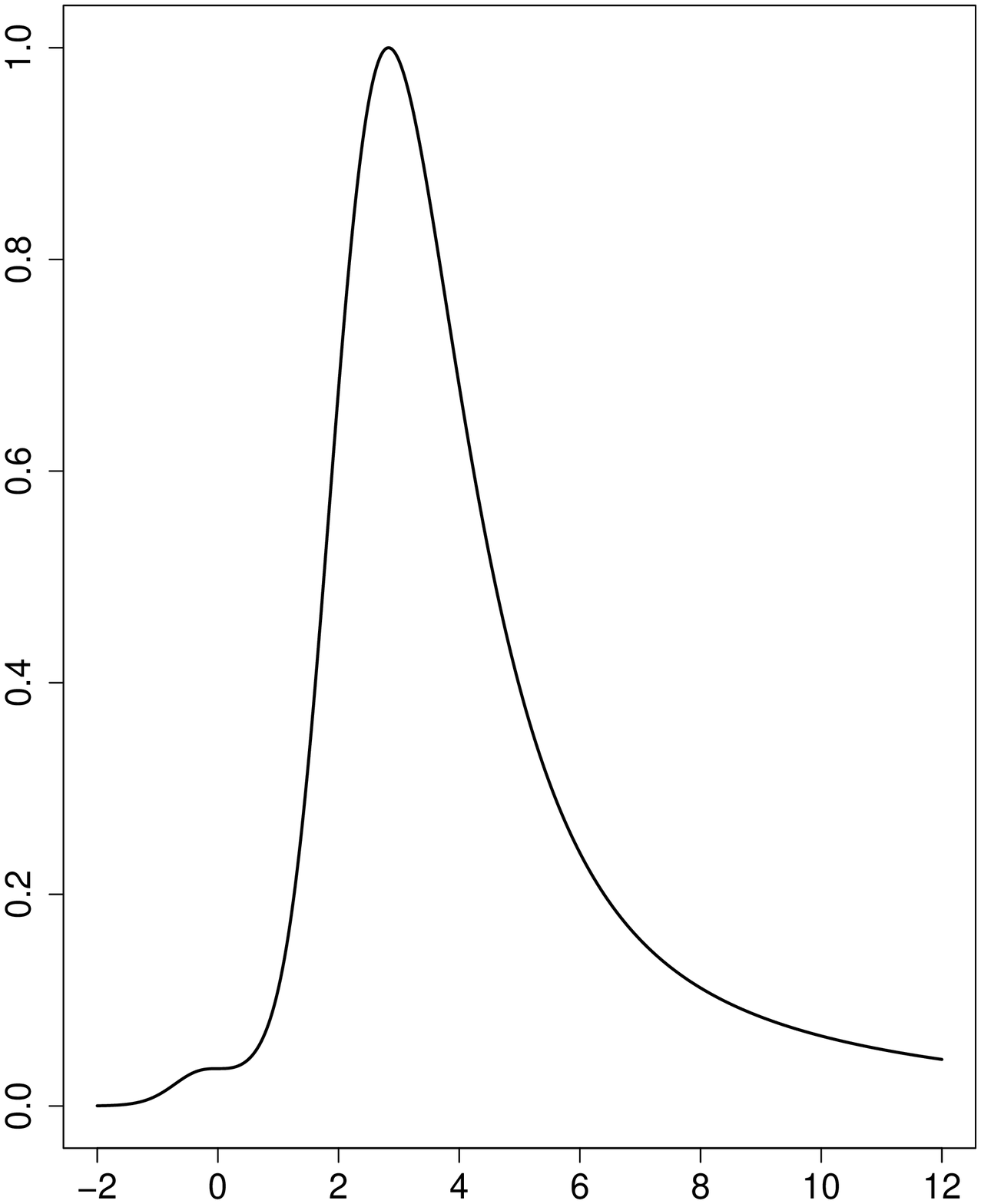}&
\includegraphics[width=5cm, height=4cm]{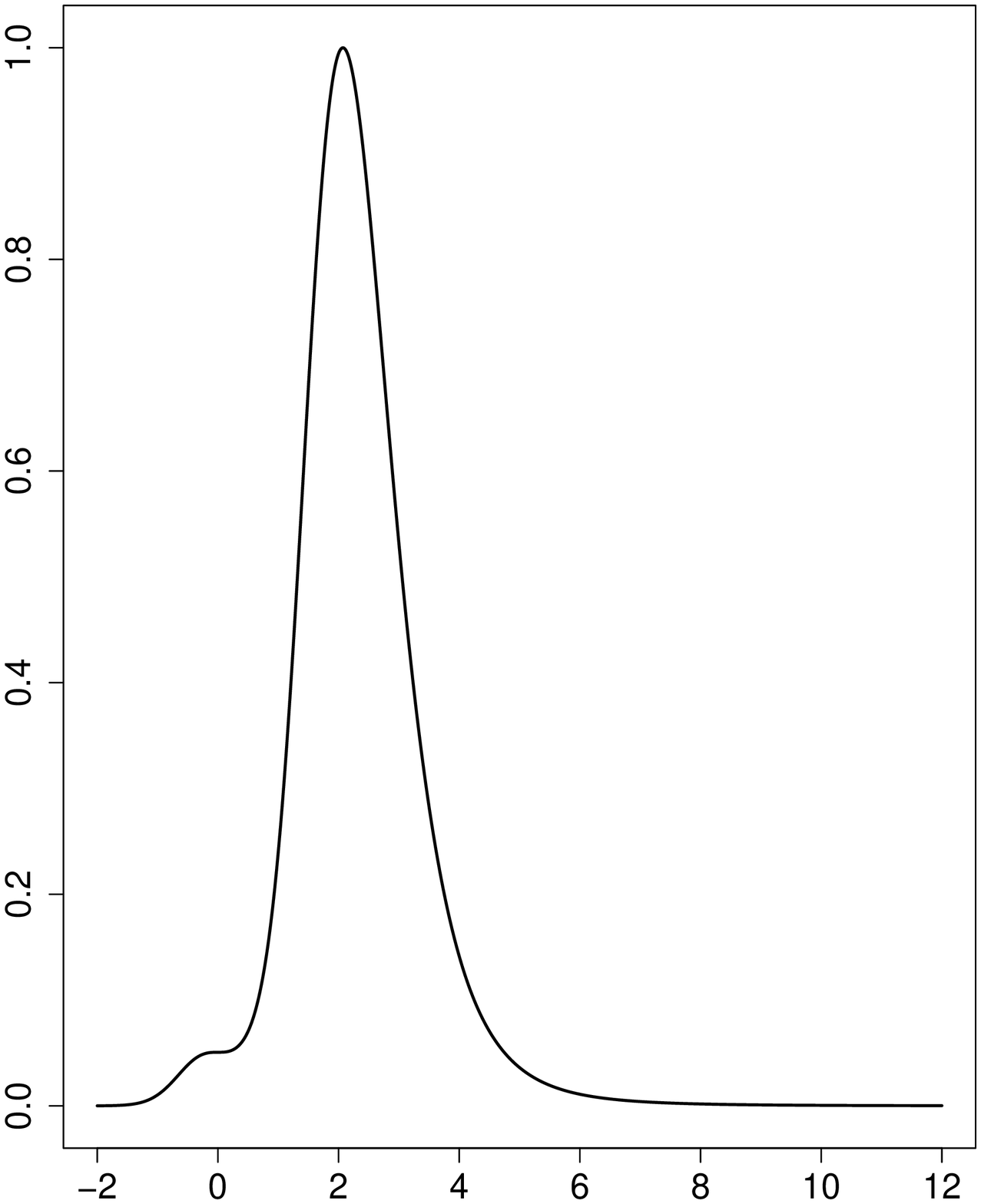}&
\includegraphics[width=5cm, height=4cm]{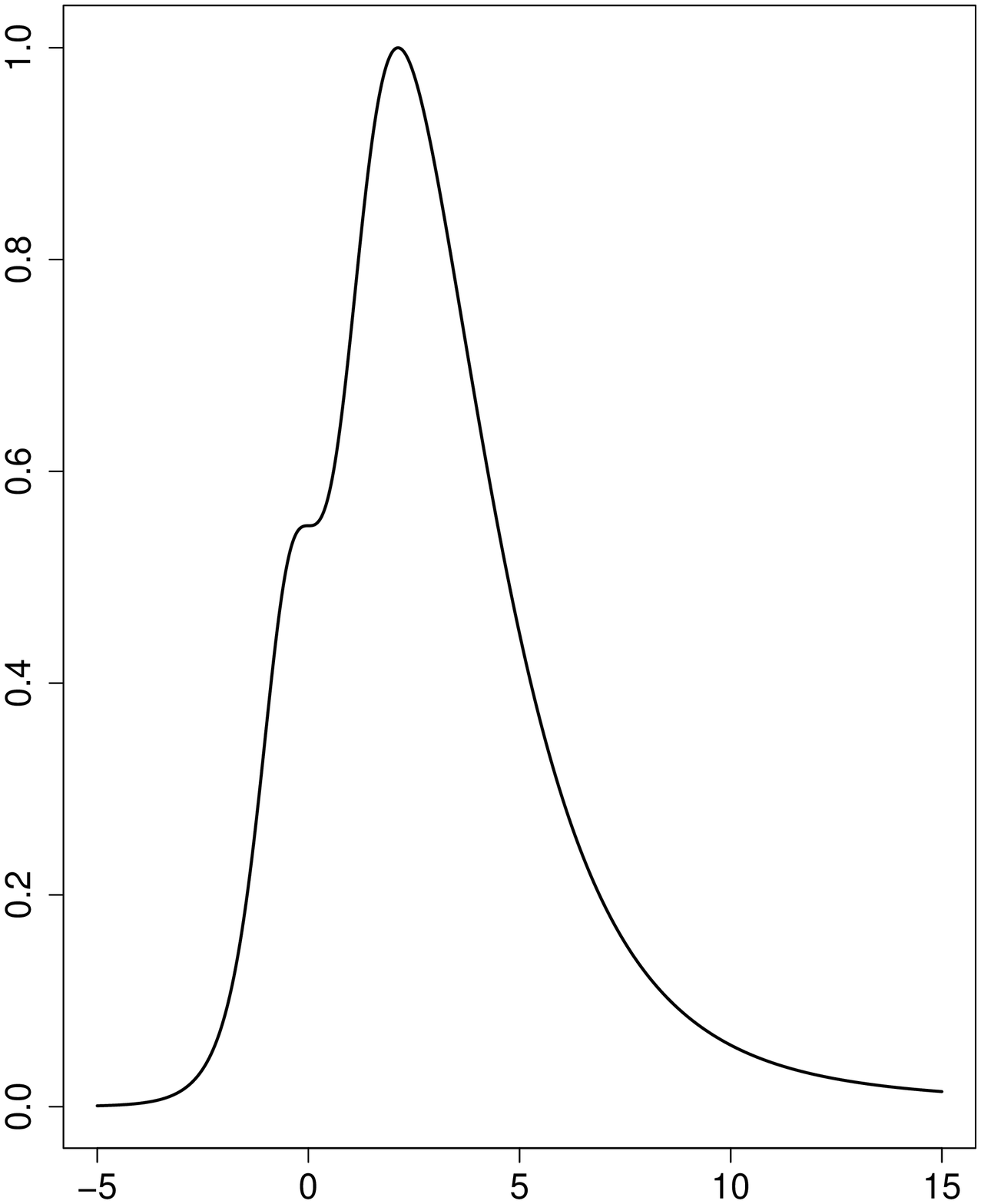}\\
(a) & (b) & (c)
\end{tabular}
\end{center}
\caption{ (a) Profile likelihood of $\lambda$ (group 18--24); (b) Profile likelihood of $\lambda$ (group 25--34); (c) Profile likelihood of $\lambda$ (group 35--44).}
\label{fig:PPLambda}
\end{figure}

For $p(\lambda)$ we use the following priors: (i) the $BTV(1,1)$ prior (\ref{PriorDI}), (ii) the $BTV(1/2,1/2)$ prior (\ref{BetaTV}), (iii) the Jeffreys prior, (iv) the informative $BTV(15.6,4.8)$ prior, (v) the informative skew-normal prior of \cite{CS13} with hyperparameters $(\mu_0,\sigma_0,\lambda_0)=(0.5,1.1,3.5)$ (which are selected in order to resemble the $BTV(15.6,4.8)$ prior), and (vi) the matching prior of  \cite{C12}. The informative prior (iv) is selected by choosing the $5\%$ and $95\%$ quantiles of the beta distribution on $M_{TV}(\lambda)$ to be $0.1$ and $0.4$, according to the expert opinion. This means favouring between $20\%$ and $80\%$ of relocation of mass to the right-hand side of the symmetry point of the baseline distribution. For each of these models, we simulate, using an adaptive MCMC sampler, a posterior sample of size $N=10,000$ from $(\mu,\sigma,\lambda)$ (with a burn-in period of $10,000$ iterations and a thinning period of $100$ iterations). Table \ref{table:BMI} shows a summary of the posterior simulations, maximum likelihood estimator of the parameters and the 95\% quantile bootstrap-confidence intervals, and the Bayes factors associated to the hypothesis $H_0: \lambda=0$ (obtained using the Savage-Dickey density ratio for priors (i)--(v), and a Laplace approximation for prior (vi)). The posterior inference for $\mu$ and $\sigma$ is similar throughout the different Bayesian models. However, we observe significant differences with respect to $\lambda$.  The credible intervals obtained with the Jeffreys, $BTV(1,1)$ and $BTV(1/2,1/2)$ priors contain negative values of $\lambda$. The reasons are that (i) these priors assign half of the mass on negative values on $\lambda$, a feature that is further increased by   their heavy tails and the flatness of the profile likelihood of $\lambda$, and (ii) the second critical point at $\lambda=0$ (see Figure \ref{fig:PPLambda}). The Matching prior induces credible intervals that do not contain the value $\lambda=0$ due to its bimodality, which pushes the credible intervals away from zero, even in cases where the distribution of the data is close to symmetry (group 35--44). These four noninformative priors include high values of $\lambda$ due to the combination of their heavy tails and the flatness of the likelihood surface in the direction of $\lambda$. On the other hand, the informative priors are centred around values which are coherent with the expert prior knowledge, in particular they only contain positive values of $\lambda$. This shows the effectiveness of working with informative priors, and the attractiveness of our new intuitive approach.

\begin{table}[!h]
\begin{center}
\begin{tabular}[h]{|c c c c c|}
\hline
Prior & $\mu$ & $\sigma$  & $\lambda$ & BF\\
\hline
\multicolumn{5}{|c|}{Age group: 18--24}\\
\hline
Jeffreys         & 21.02 (19.67,23.96) & 3.88 (2.52,4.92)  & 2.57 (-1.11,5.97)  &  0.31 \\
BTV(1/2,1/2)     & 21.05 (19.71,24.04) & 3.84 (2.55,4.95)  & 2.48 (-0.85,6.09)  &  0.27  \\
BTV(1,1)         & 21.17 (19.82,24.60) & 3.75 (2.50,4.87)  & 2.24 (-0.78,5.69)  &  0.28  \\
BTV(15.6,4.8)    & 21.33 (20.20,22.61) & 3.61 (2.69,4.56)  & 1.94 ( 0.50,3.73)  &  0.16  \\
SN               & 21.33 (20.39,22.54) & 3.62 (2.79,4.47)  & 1.94 ( 0.59,3.23)  &  0.15 \\
Matching         & 20.88 (19.73,21.96) & 3.99 (2.98,5.04)  & 2.90 ( 0.84,7.17)  & 0.23  \\
MLE              & 20.94 (19.73,21.87) & 3.93 (2.94,4.97)  &  2.83 ( 1.47,$\infty$) & --  \\
\hline
\multicolumn{5}{|c|}{Age group: 25--34}\\
\hline
Jeffreys         &  21.73 (20.20,25.02) & 4.16 (2.89,5.23)  & 1.80 (-0.43,3.70) &  0.45  \\
BTV(1/2,1/2)     &  21.73 (20.16,25.00) & 4.14 (2.92,5.29)  & 1.79 (-0.63,3.51) &  0.44 \\
BTV(1,1)         &  21.85 (20.34,25.37) & 4.05 (2.86,5.17)  & 1.64 (-0.62,3.38) &  0.36  \\
BTV(15.6,4.8)    &  21.83 (20.67,23.25) & 4.06 (3.19,5.13)  & 1.64 ( 0.51,2.98) &  0.21  \\
SN               &  21.80 (20.74,23.18) & 4.10 (3.17,5.04)  & 1.70 ( 0.63,2.92) &  0.23  \\
Matching         & 21.47 (20.26,22.81) & 4.36 (3.31,5.38)  & 2.11 ( 0.78,4.13)  & 0.39  \\
MLE              &  21.53 (20.53,24.32) & 4.29 (2.70,5.37)  & 2.07 ( 0.02,3.87) & --  \\
\hline
\multicolumn{5}{|c|}{Age group: 35--44}\\
\hline
Jeffreys         & 24.30 (21.53,28.33) & 3.34 (2.25,5.02)  & 0.74 (-2.07,4.87)  & 1.54  \\
BTV(1/2,1/2)     & 24.29 (21.42,28.15) & 3.31 (2.29,5.04)  & 0.75 (-1.79,5.10)  & 1.38  \\
BTV(1,1)         & 24.72 (21.71,28.21) & 3.20 (2.24,4.77)  & 0.49 (-1.96,3.87)  & 1.07  \\
BTV(15.6,4.8)    & 23.53 (21.94,25.30) & 3.48 (2.42,4.69)  & 1.32 ( 0.05,3.16)  & 0.93  \\
SN               & 23.48 (22.01,25.17) & 3.50 (2.50,4.67)  & 1.36 ( 0.23,2.88)  & 0.78  \\
Matching         & 22.98 (21.22,28.79) & 3.96 (2.73,5.61)  & 2.05 (-2.33,7.00)  &  2.63 \\
MLE              & 22.95 (21.58,26.18) & 3.84 (2.26,5.10)  & 2.12 ( 0,$\infty$) & --  \\
\hline
\end{tabular}
\caption{ BMI data: posterior median, 95\% posterior credible intervals, and Bayes factors associated to $\mathcal{H}_0:\lambda=0$.}
\label{table:BMI}
\end{center}
\end{table}

\section{Discussion}

The construction of meaningful priors, either informative or noninformative, is of central importance in Bayesian inference. Prior elicitation is particularly challenging when the model parameters control several features. Such is the case of the skewness, or rather perturbation, parameter in skew-symmetric distributions. This parameter controls the mode, asymmetry, tail behaviour, and spread of the pdf. We proposed a new method for constructing priors for this parameter based on its overall effect on the shape of the density. For this purpose, we studied the perturbation effect of the shape parameter through the Total Variation distance. We showed that the priors induced by the Total Variation distance are very intuitive and hence user-friendly, have very good frequentist properties and enjoy  tractable expressions, especially compared to the popular Jeffreys prior which moreover can have singularities.

The constructive strategy proposed in this paper can be extended to shape parameters in other distributions.  In the Supplementary Material, we provide a brief study on the construction of priors using the Total Variation distance for log-skew-symmetric distributions and two-piece distributions. It is shown that the priors for the entire family of two-piece distributions have closed-form expressions, which are linked to a family of priors proposed in \cite{RS14}. Applying this new strategy of prior construction to various other families with shape parameters  represents a promising research direction. \cite{S14} recently proposed a new strategy to construct prior distributions based on the Kullback-Leibler divergence. The reasoning behind those priors is to penalise model complexity by constructing a prior that favours the baseline nested model, which differs from the approach presented here. The study of Penalised Complexity priors for skew-symmetric models represents a possible research direction. More generally, the use of other distances between the distributions, instead of the Total Variation, is another interesting research direction, although this has to be done with some care as different distances may have very different properties. For instance, a possible competitor is the Wasserstein distance, which is defined for two distributions $F_1$ and $F_2$ on ${\mathbb R}$, with finite first moment, by \citep{V74}
\begin{eqnarray*}
d_{\mathcal W}(F_1,F_2) = \int_{\mathbb R} \vert F_1(x)-F_2(x)\vert dx.
\end{eqnarray*}
Unfortunately, this distance is not invariant under monotone transformations, a property used in Section \ref{sec:distances} to construct a measure of perturbation. Other distances, such as the Energy distance, Kolmogorov distance, and the Hellinger distance, do not seem to lead to closed-form or interpretable expressions, in contrast to our choice.

The priors proposed in this paper can be extended to the family of multivariate skew-symmetric distributions with density function:
\begin{eqnarray*}
s_n({\bf x}; \bm{\lambda}) = 2f_n({\bf x})G(\bm{\lambda}\cdot{\bf x}),
\end{eqnarray*}
where ${\bf x} = (x_1,\dots,x_n)\in{\mathbb R}^n$, $\bm{\lambda}=(\lambda_1,\dots,\lambda_n)\in{\mathbb R}^n$, $f_n$ is a symmetric pdf with support on ${\mathbb R}^n$, $G$ is a symmetric cdf with support on ${\mathbb R}$, and $\bm{\lambda}\cdot{\bf x}= \sum_{j=1}^n \lambda_j x_j$. In order to construct a joint prior for the skewness parameter $\bm{\lambda}$, we can decompose it into conditional priors:
\begin{eqnarray*}\label{MultPrior}
\pi(\bm{\lambda}) = \pi(\lambda_1\vert \lambda_2,\dots,\lambda_n)\pi(\lambda_2\vert \lambda_3,\dots,\lambda_n)\dots \pi(\lambda_n),
\end{eqnarray*}
and sequentially apply the strategy proposed in Section \ref{ProposedPriorSect} to construct the univariate conditional priors. A more detailed study of this prior will be considered for future research.

\section*{Appendix: Proofs} \label{proof}
\subsection*{Proof of the representation  \eqref{MTVSMSN} for skew-$t$ distributions}
Let $X_{\lambda,\nu}$ be a random variable following a skew-$t$ distribution. By using the stochastic representation of the skew-$t$ distribution as a scale mixture of skew-normal distributions \citep{AC03} it follows that
\begin{eqnarray}\label{SMSN}
S_{t_\nu,T_{\nu+1}}(0;\lambda,\nu) &=& {\mathbb P}\left(X_{\lambda,\nu} \leq 0\right) ={\mathbb P}\left(V_{\nu}^{-1/2}Z_{\lambda} \leq 0\right) ={\mathbb P}\left(Z_\lambda \leq 0\right),
\end{eqnarray}
where $V_{\nu}\sim \chi^2_{\nu}/\nu$, and $Z_{\lambda}$ is a skew-normal random variable with location $0$, unit scale, and skewness parameter $\lambda$. The result follows from this relationship together with equation \eqref{MTVClosedForm}.

We emphasize that not all  scale mixtures of skew-normal distributions can be written as a skew-symmetric distribution of the type \eqref{sspdf}. The skew-$t$ is one of few cases reported in the literature which admit both representations. However, equation \eqref{SMSN} reveals a more general property of scale mixtures of skew-normal distributions since ${\mathbb P}\left(V^{-1/2}Z_{\lambda} \leq 0\right) ={\mathbb P}\left(Z_\lambda \leq 0\right)$ for any positive random variable $V^{-1/2}$.

\subsection*{Proof of Theorem \ref{CharactTV}}
\begin{enumerate}[(i)]
\item The symmetry property is immediate from expression  (\ref{PriorDI}).
\item It is easily seen that $\pi_{TV}(\lambda)  = \int_0^{\infty} 2\omega(u) f(u)g(\lambda \omega(u))du.$
For $u>0$ and $|\lambda_1|>|\lambda_2|>0$, it follows that $\omega(u) f(u)g(\lambda_1 \omega(u)) < \omega(u) f(u)g(\lambda_2 \omega(u))$ thanks to the unimodality and symmetry of $g$. Thus $$
\int_0^{\infty} \omega(u) f(u)g(\lambda_1 \omega(u))du \leq \int_0^{\infty} \omega(u) f(u)g(\lambda_2 \omega(u))du$$
and hence the prior is decreasing in $|\lambda|$.
\item By using the change of variable $u = \lambda x$ and the maximality of $f$ at 0, it follows that for $\lambda>0$
\begin{eqnarray*}
\int_0^{\infty} x f(x)g(\lambda x)dx   \leq M\int_0^{\infty} x g(\lambda x)dx = \dfrac{M}{\lambda^2}\int_0^{\infty} u g(u)du.
\end{eqnarray*}
The same results hold true for $\lambda<0$ by the symmetry property shown in (i). Now, let $|\lambda|\geq L>0$. Then, the unimodality and symmetry of $f$ yield  $f\left(\dfrac{x}{\lambda}\right)\geq f\left(\dfrac{x}{L}\right)$ for $x>0$. By using the change of variable $u=\lambda x$ we find that for $\lambda>0$
\begin{eqnarray*}
 \int_0^{\infty} x f(x)g(\lambda x)dx  &=&   \dfrac{1}{\lambda^2}\int_0^{\infty} u f\left(\dfrac{u}{\lambda}\right) g(u)du \\
 &\geq& \dfrac{1}{\lambda^2} \int_0^{\infty} u f\left(\dfrac{u}{L}\right)g(u)du.
\end{eqnarray*}
Analogously for $\lambda<0$. The result follows by combining the previous inequalities.
\end{enumerate}
\subsection*{Proof of Theorem \ref{PointObs}}

 Recall that a posterior distribution is proper whenever the marginal distribution $P(x_1,\dots,x_n)<\infty$ \citep{FS99}. Now note that
$
s_{f;G}(x;\mu,\sigma,\lambda) \leq \dfrac{2}{\sigma}f\left(\dfrac{x-\mu}{\sigma}\right),
$
 which entails that
\begin{eqnarray*}
P(x_1,\dots,x_n) &=& \int_{\mathbb R} \int_{{\mathbb R}_+} \int_{\mathbb R}  \left[\prod_{j=1}^n s(x_j;\mu,\sigma,\lambda)\right] \dfrac{p(\lambda)}{\sigma}d\mu d\sigma d\lambda\\
&\leq& \int_{{\mathbb R}_+} \int_{\mathbb R} \left[\prod_{j=1}^n \dfrac{2}{\sigma}f\left(\dfrac{x_j-\mu}{\sigma}\right)\right]\dfrac{1}{\sigma}d\mu d\sigma \int_{\mathbb R} p(\lambda)d\lambda.
\end{eqnarray*}
Given that $p(\lambda)$ is proper, it follows that the posterior distribution of $(\mu,\sigma,\lambda)$ exists whenever the posterior distribution of $(\mu,\sigma)$ exists for a scale mixture of normals sampling model and the prior $\pi(\mu,\sigma)\propto \sigma^{-1}$. The propriety of the latter, for $n\geq 2$ and when all the observations are different, follows by Theorem 1 of \cite{FS99}.

\section*{Acknowledgements}
Holger Dette's work has  been supported in part by the Collaborative Research Center ``Statistical modeling of nonlinear dynamic processes'' (SFB 823, Teilprojekt C1) of the German Research Foundation (DFG). The authors thank Helena Carreira (LSHTM) for helpful discussions on the behaviour of BMI data, and two anonymous referees for useful comments.

\newpage
\section*{Supplementary Material: Extensions and Monte Carlo simulation study}

\section{Extension to other distributions}

We here briefly show how our method applies to other types of flexible distributions where a (skewness) parameter has a perturbation effect on the original distribution.

\subsection*{Log-skew-symmetric distributions}
The proposed priors from our main paper have the same interpretation if they are used for the perturbation parameter in log-skew-symmetric distributions. Recall that a positive random variable $Y$ is said to be distributed according to a log-skew-symmetric distribution if it is distributed according to (\ref{LSSPDF}) below. This sort of distributions have been used for modelling environmental, medical, biological, and financial data (see \citealp{MG10} and the references therein). The pdf of $Y$ is given by
\begin{eqnarray}\label{LSSPDF}
s_l(y;\lambda) = \dfrac{2}{y}f(\log y)G(\lambda \log y), \,\,\, y>0.
\end{eqnarray}
It follows that the TV distance between $(\ref{LSSPDF})$ and the corresponding baseline log-symmetric density $f_l(y)=\dfrac{1}{y}f(\log y)$ satisfies
\begin{eqnarray*}
d_{TV}(s_l,f_l\vert\lambda) &=& \dfrac{1}{2}\int_0^{\infty} \left\vert  s_l(y;\lambda) - f_l(y) \right\vert dy =  \dfrac{1}{2}\int_{-\infty}^{\infty} \vert  s(x;\lambda) - f(x) \vert dx = d_{TV}(s,f\vert\lambda).
\end{eqnarray*}
Consequently, the priors proposed in Section~3 of the main paper for the skew-symmetric family coincide with those obtained for the log-skew-symmetric family. It is also clear that one could use any other increasing diffeomorphism from ${\mathbb R}_+$  to ${\mathbb R}$ instead of the logarithmic transformation.

\subsection*{Two-piece distributions}
Consider the family of two-piece distributions with the following parameterisation (see \citealp{RS14} for a general overview):
\begin{eqnarray}\label{TPPDF}
s_{tp}(x;\gamma) = f\left(\dfrac{x}{1-\gamma}\right)I(x<0) +  f\left(\dfrac{x}{1+\gamma}\right)I(x\geq 0),\,\,\,x\in{\mathbb R},
\end{eqnarray}
where $\gamma \in(-1,1)$, and $f$ is a unimodal symmetric pdf with mode at $0$. The parameter $\gamma$ controls the mass cumulated on either side of the mode ($x=0$) while preserving the tail behaviour of~$f$. Density (\ref{TPPDF}) is asymmetric for $\gamma\neq0$ and it reduces to $f$ for $\gamma=0$. The TV distance between $s_{tp}$ and the baseline pdf $f$ is given by:
\begin{eqnarray*}
d_{TV}(s_{tp},f\vert\gamma) &=&
\dfrac{1}{2}\int_{-\infty}^0 \left| f\left(\dfrac{x}{1-\gamma}\right)-f(x)\right| dx + \dfrac{1}{2}\int_0^{\infty} \left|f\left(\dfrac{x}{1+\gamma}\right)  -f(x)\right|dx = \dfrac{\vert \gamma \vert}{2}.
\end{eqnarray*}
If we define the measure of perturbation $M_{TV}(\gamma) = \gamma/2$, this coincides, up to a proportionality constant, with the AG measure of skewness proposed in \cite{AG95} (see \citealp{RS14}). Consequently, if we assume that $M_{TV}(\gamma) = \gamma/2 \sim Beta(\alpha,\beta)$ (using the notation in the main paper for the Beta distribution), we obtain the AG-Beta priors proposed in \cite{RS14} for this family of distributions.

\section{Simulation study}\label{SimStudySectS}

This section is a complement to the Monte Carlo simulation study in Section~4.1 of the main paper. It is a performance comparison between noninformative priors built according to our new method and the Jeffreys prior.

We simulate $N=1,000$ samples of sizes $n=100$ and $n=200$ from the skew-normal, skew-logistic and skew-Laplace distributions,  in each case with location parameter $\mu=0$, scale parameter $\sigma=1$, and skewness parameter $\lambda=0,2.5,5$. For each of these samples, we simulate a posterior sample of size $1,000$ from $(\mu,\sigma,\lambda)$ using the  $BTV(1,1)$, $BTV(1/2,1/2)$, and Jeffreys priors. We employ a self-adaptive MCMC sampler  to obtain the posterior samples. For each posterior sample, we calculate the coverage proportions of the 95\% credible intervals of each parameter (that is, the proportion of credible intervals that contain the true value of the parameter) as well as the 5\%, 50\% and 95\% quantiles of the posterior medians and maximum \emph{a posteriori} (MAP) estimators. In addition, we obtain the median of the Bayes factors (BFs) associated to the hypothesis $H_0:\lambda=0$. The Bayes factors are approximated using the Savage-Dickey density ratio. Results are reported in Tables \ref{table:SN100}--\ref{table:SL200}. Overall, we observe that the Jeffreys and $BTV(1/2,1/2)$ priors exhibit the best, and very similar, performance.



\begin{table}[!htbp]
\begin{center}
\scriptsize
\begin{tabular}[h]{|c| c c c| c c c| c| c|}
\hline
Prior &  \multicolumn{3}{c}{MAP} & \multicolumn{3}{c}{Median} & Coverage & BF \\
\hline
 & 5\% & 50\% & 95\% & 5\% & 50\% & 95\% & &\\
  \hline
$\lambda=0$&&&&&&&&\\
\hline
BTV(1/2,1/2) &&&&&&&&\\
$\mu$ & -1.031 & 0.018 & 0.993 & -0.797 & 0.009 & 0.775 & 0.983 & --\\
  $\sigma$ & 0.943 & 1.074 & 1.389 & 1.000 & 1.150 & 1.387 & 0.861 & --\\
  $\lambda$ & -1.455 & 0.001 & 1.481 & -1.244 & -0.018 & 1.315 & 0.984 &  2.002\\
Jeffreys &&&&&&&&\\
$\mu$ & -1.039 & 0.028 & 1.020 & -0.864 & 0.011 & 0.800 & 0.982 & --\\
  $\sigma$ & 0.946 & 1.084 & 1.432 & 1.008 & 1.159 & 1.400 & 0.839 & --\\
  $\lambda$ & -1.537 & -0.021 & 1.482 & -1.392 & -0.026 & 1.401 & 0.982 &  2.204\\
BTV(1,1) &&&&&&&&\\
$\mu$ & -0.997 & -0.007 & 0.960 & -0.708 & 0.015 & 0.683 & 0.992 & --\\
  $\sigma$ & 0.933 & 1.067 & 1.355 & 0.985 & 1.131 & 1.346 & 0.877 & --\\
  $\lambda$ & -1.186 & 0.008 & 1.279 & -0.992 & -0.010 & 1.075 & 0.992 & 1.404\\
\hline
$\lambda=2.5$&&&&&&&&\\
\hline
BTV(1/2,1/2) &&&&&&&&\\
$\mu$ & -0.188 & 0.014 & 0.646 & -0.170 & 0.058 & 0.635 & 0.880 & --\\
  $\sigma$ & 0.650 & 0.955 & 1.177 & 0.704 & 0.947 & 1.175 & 0.898 & --\\
  $\lambda$ & -0.029 & 2.111 & 4.318 & 0.212 & 2.139 & 5.127 & 0.879 & 0.386\\
Jeffreys &&&&&&&&\\
$\mu$ & -0.202 & 0.009 & 0.478 & -0.174 & 0.052 & 0.638 & 0.890 & --\\
  $\sigma$ & 0.651 & 0.959 & 1.180 & 0.708 & 0.952 & 1.177 & 0.919 & --\\
  $\lambda$ & 0.003 & 2.138 & 4.308 & 0.274 & 2.191 & 5.028 & 0.885 & 0.369\\
BTV(1,1) &&&&&&&&\\
$\mu$ & -0.176 & 0.031 & 0.739 & -0.148 & 0.088 & 0.667 & 0.870 & --\\
  $\sigma$ & 0.645 & 0.931 & 1.155 & 0.691 & 0.922 & 1.150 & 0.897 & --\\
  $\lambda$ & -0.069 & 1.955 & 3.867 & 0.151 & 1.950 & 4.504 & 0.859 & 0.338\\
\hline
$\lambda=5$&&&&&&&&\\
\hline
BTV(1/2,1/2) &&&&&&&&\\
$\mu$ & -0.142 & -0.006 & 0.152 & -0.120 & 0.004 & 0.185 & 0.922 & --\\
  $\sigma$ & 0.806 & 0.983 & 1.137 & 0.814 & 0.988 & 1.147 & 0.946 & --\\
  $\lambda$ & 1.724 & 4.245 & 10.966 & 2.289 & 4.995 & 14.257 & 0.931 & 0.004\\
Jeffreys &&&&&&&&\\
$\mu$ & -0.134 & -0.003 & 0.150 & -0.123 & 0.004 & 0.184 & 0.925 & --\\
  $\sigma$ & 0.810 & 0.986 & 1.147 & 0.817 & 0.991 & 1.147 & 0.944 & --\\
  $\lambda$ & 1.893 & 4.336 & 10.033 & 2.409 & 5.044 & 15.041 & 0.927 & 0.004\\
BTV(1,1) &&&&&&&&\\
$\mu$ & -0.117 & 0.011 & 0.177 & -0.103 & 0.019 & 0.231 & 0.922 & --\\
  $\sigma$ & 0.788 & 0.975 & 1.133 & 0.780 & 0.975 & 1.131 & 0.935 & --\\
  $\lambda$ & 1.767 & 4.044 & 8.088 & 1.945 & 4.582 & 10.286 & 0.932 & 0.003\\
\hline
\end{tabular}
\caption{Skew-normal data for noninformative priors: $\mu=0,\sigma=1, n=100$.}
\label{table:SN100}
\end{center}
\end{table}

\begin{table}[!htbp]
\begin{center}
\scriptsize
\begin{tabular}[h]{|c| c c c| c c c| c| c|}
\hline
Prior &  \multicolumn{3}{c}{MAP} & \multicolumn{3}{c}{Median} & Coverage & BF \\
\hline
 & 5\% & 50\% & 95\% & 5\% & 50\% & 95\% & &\\
  \hline
$\lambda=0$&&&&&&&&\\
\hline
BTV(1/2,1/2) &&&&&&&&\\
$\mu$ & -0.886 & -0.035 & 0.869 & -0.682 & -0.009 & 0.692 & 0.994 & --\\
  $\sigma$ & 0.960 & 1.054 & 1.330 & 1.008 & 1.108 & 1.296 & 0.846 & --\\
  $\lambda$ & -1.286 & 0.028 & 1.295 & -1.028 & -0.004 & 1.037 & 0.993 & 2.306\\
Jeffreys &&&&&&&&\\
$\mu$ & -0.890 & 0.032 & 0.890 & -0.718 & 0.006 & 0.720 & 0.991 & --\\
  $\sigma$ & 0.961 & 1.058 & 1.351 & 1.009 & 1.115 & 1.319 & 0.844 & --\\
  $\lambda$ & -1.335 & -0.002 & 1.297 & -1.082 & -0.019 & 1.052 & 0.991 &  2.557\\
BTV(1,1) &&&&&&&&\\
$\mu$ & -0.857 & 0.002 & 0.860 & -0.602 & -0.007 & 0.652 & 0.995 & --\\
  $\sigma$ & 0.952 & 1.048 & 1.285 & 0.999 & 1.098 & 1.275 & 0.862 & --\\
  $\lambda$ & -1.207 & -0.001 & 1.180 & -0.992 & -0.004 & 0.851 & 0.995 & 1.570\\
\hline
$\lambda=2.5$&&&&&&&&\\
\hline
BTV(1/2,1/2) &&&&&&&&\\
$\mu$ & -0.124 & 0.008 & 0.211 & -0.119 & 0.021 & 0.317 & 0.905 & --\\
  $\sigma$ & 0.699 & 0.981 & 1.123 & 0.760 & 0.980 & 1.124 & 0.909 & --\\
  $\lambda$ & 1.158 & 2.283 & 3.671 & 0.924 & 2.362 & 3.814 & 0.907 & 0.012\\
Jeffreys &&&&&&&&\\
$\mu$ & -0.127 & 0.006 & 0.202 & -0.119 & 0.019 & 0.321 & 0.914 & --\\
  $\sigma$ & 0.709 & 0.984 & 1.120 & 0.772 & 0.982 & 1.125 & 0.911 & --\\
  $\lambda$ & 1.191 & 2.319 & 3.645 & 0.922 & 2.369 & 3.886 & 0.916 & 0.006\\
BTV(1,1) &&&&&&&&\\
$\mu$ & -0.123 & 0.015 & 0.225 & -0.110 & 0.031 & 0.401 & 0.898 & --\\
  $\sigma$ & 0.687 & 0.976 & 1.123 & 0.743 & 0.972 & 1.117 & 0.902 & --\\
  $\lambda$ & 0.787 & 2.231 & 3.541 & 0.722 & 2.294 & 3.666 & 0.897 & 0.014\\
\hline
$\lambda=5$&&&&&&&&\\
\hline
BTV(1/2,1/2) &&&&&&&&\\
$\mu$ & -0.090 & 0.000 & 0.097 & -0.082 & 0.003 & 0.098 & 0.938 & --\\
  $\sigma$ & 0.874 & 0.992 & 1.108 & 0.874 & 0.995 & 1.107 & 0.935 & --\\
  $\lambda$ & 2.948 & 4.603 & 7.795 & 3.129 & 4.939 & 8.662 & 0.923 & $5\times10^{-11}$\\
Jeffreys &&&&&&&&\\
$\mu$ & -0.087 & -0.001 & 0.095 & -0.081 & 0.004 & 0.098 & 0.940 & --\\
  $\sigma$ & 0.873 & 0.989 & 1.099 & 0.874 & 0.995 & 1.108 & 0.937 & --\\
  $\lambda$ & 2.922 & 4.615 & 7.907 & 3.134 & 4.947 & 8.943 & 0.926 & $4\times10^{-11}$\\
BTV(1,1) &&&&&&&&\\
$\mu$ & -0.080 & 0.004 & 0.098 & -0.076 & 0.008 & 0.106 & 0.945 & --\\
  $\sigma$ & 0.867 & 0.987 & 1.096 & 0.869 & 0.990 & 1.101 & 0.944 & --\\
  $\lambda$ & 2.853 & 4.457 & 7.475 & 3.048 & 4.748 & 8.275 & 0.929 & $5\times10^{-11}$\\
\hline
\end{tabular}
\caption{Skew-normal data for noninformative priors: $\mu=0,\sigma=1, n=200$.}
\label{table:SN200}
\end{center}
\end{table}

\begin{table}[!htbp]
\begin{center}
\scriptsize
\begin{tabular}[h]{|c| c c c| c c c| c| c|}
\hline
Prior &  \multicolumn{3}{c}{MAP} & \multicolumn{3}{c}{Median} & Coverage & BF \\
\hline
 & 5\% & 50\% & 95\% & 5\% & 50\% & 95\% & &\\
  \hline
$\lambda=0$&&&&&&&&\\
\hline
BTV(1/2,1/2) &&&&&&&&\\
$\mu$ & -1.139 & 0.048 & 1.191 & -0.994 & 0.055 & 1.046 & 0.957 & --\\
  $\sigma$ & 0.906 & 1.060 & 1.244 & 0.936 & 1.099 & 1.282 & 0.911 & --\\
  $\lambda$ & -0.682 & -0.014 & 0.641 & -0.842 & -0.030 & 0.820 & 0.952 & 2.794\\
Jeffreys &&&&&&&&\\
$\mu$ & -1.190 & 0.047 & 1.227 & -1.044 & 0.059 & 1.097 & 0.952 & --\\
  $\sigma$ & 0.913 & 1.070 & 1.254 & 0.944 & 1.106 & 1.295 & 0.910 & --\\
  $\lambda$ & -0.783 & -0.014 & 0.719 & -0.899 & -0.044 & 0.865 & 0.946 & 2.913\\
BTV(1,1) &&&&&&&&\\
$\mu$ & -1.030 & 0.029 & 1.082 & -0.932 & 0.051 & 0.965 & 0.965 & --\\
  $\sigma$ & 0.901 & 1.054 & 1.230 & 0.938 & 1.092 & 1.270 & 0.916 & --\\
  $\lambda$ & -0.614 & -0.016 & 0.608 & -0.766 & -0.035 & 0.739 & 0.961 & 1.901\\
\hline
$\lambda=2.5$&&&&&&&&\\
\hline
BTV(1/2,1/2) &&&&&&&&\\
$\mu$ & -0.312 & 0.037 & 0.546 & -0.278 & 0.090 & 0.602 & 0.922 & --\\
  $\sigma$ & 0.706 & 0.947 & 1.196 & 0.734 & 0.950 & 1.203 & 0.923 & --\\
  $\lambda$ & 0.540 & 1.929 & 4.478 & 0.766 & 2.141 & 5.002 & 0.904 & 0.122\\
Jeffreys &&&&&&&&\\
$\mu$ & -0.305 & 0.031 & 0.535 & -0.277 & 0.089 & 0.604 & 0.925 & --\\
  $\sigma$ & 0.704 & 0.949 & 1.203 & 0.737 & 0.955 & 1.204 & 0.921 & --\\
  $\lambda$ & 0.565 & 1.951 & 4.205 & 0.742 & 2.137 & 5.028 & 0.911 & 0.115\\
BTV(1,1) &&&&&&&&\\
$\mu$ & -0.264 & 0.074 & 0.642 & -0.237 & 0.134 & 0.676 & 0.918 & --\\
  $\sigma$ & 0.685 & 0.917 & 1.174 & 0.718 & 0.926 & 1.180 & 0.913 & --\\
  $\lambda$ & 0.496 & 1.725 & 3.919 & 0.675 & 1.917 & 4.491 & 0.894 & 0.099\\
\hline
$\lambda=5$&&&&&&&&\\
\hline
BTV(1/2,1/2) &&&&&&&&\\
$\mu$ & -0.222 & -0.003 & 0.260 & -0.198 & 0.018 & 0.294 & 0.931 & --\\
  $\sigma$ & 0.780 & 0.973 & 1.168 & 0.792 & 0.980 & 1.175 & 0.942 & --\\
  $\lambda$ & 1.649 & 4.077 & 10.223 & 2.174 & 4.733 & 13.117 & 0.916 & 0.006\\
Jeffreys &&&&&&&&\\
$\mu$ & -0.210 & 0.001 & 0.264 & -0.200 & 0.015 & 0.287 & 0.933 & --\\
  $\sigma$ & 0.788 & 0.975 & 1.168 & 0.796 & 0.981 & 1.172 & 0.943 & --\\
  $\lambda$ & 1.507 & 4.109 & 9.910 & 2.178 & 4.766 & 13.401 & 0.907& 0.006 \\
BTV(1,1) &&&&&&&&\\
$\mu$ & -0.182 & 0.020 & 0.303 & -0.170 & 0.040 & 0.330 & 0.930 & --\\
  $\sigma$ & 0.761 & 0.961 & 1.156 & 0.774 & 0.967 & 1.155 & 0.932 & --\\
  $\lambda$ & 1.601 & 3.792 & 7.933 & 1.964 & 4.338 & 10.000 & 0.909 & 0.005\\
\hline
\end{tabular}
\caption{ Skew-logistic data for noninformative priors: $\mu=0,\sigma=1, n=100$.}
\label{table:SLO100}
\end{center}
\end{table}

\begin{table}[!htbp]
\begin{center}
\scriptsize
\begin{tabular}[h]{|c| c c c| c c c| c| c|}
\hline
Prior &  \multicolumn{3}{c}{MAP} & \multicolumn{3}{c}{Median} & Coverage & BF \\
\hline
 & 5\% & 50\% & 95\% & 5\% & 50\% & 95\% & &\\
  \hline
$\lambda=0$&&&&&&&&\\
\hline
BTV(1/2,1/2) &&&&&&&&\\
$\mu$ & -0.907 & -0.022 & 0.904 & -0.824 & -0.010 & 0.807 & 0.938 & --\\
  $\sigma$ & 0.931 & 1.037 & 1.168 & 0.950 & 1.059 & 1.187 & 0.913 & --\\
  $\lambda$ & -0.529 & -0.002 & 0.540 & -0.594 & -0.002 & 0.632 & 0.938 & 3.838\\
Jeffreys &&&&&&&&\\
$\mu$ & -0.913 & -0.012 & 0.902 & -0.826 & -0.007 & 0.820 & 0.936 & --\\
  $\sigma$ & 0.932 & 1.038 & 1.165 & 0.949 & 1.061 & 1.191 & 0.915 & --\\
  $\lambda$ & -0.531 & 0.011 & 0.566 & -0.606 & 0.002 & 0.662 & 0.931 & 4.186\\
BTV(1,1) &&&&&&&&\\
$\mu$ & -0.879 & -0.013 & 0.837 & -0.780 & -0.012 & 0.755 & 0.945 & --\\
  $\sigma$ & 0.926 & 1.035 & 1.154 & 0.946 & 1.058 & 1.181 & 0.921 & --\\
  $\lambda$ & -0.481 & 0.007 & 0.542 & -0.531 & 0.003 & 0.618 & 0.942 & 2.574\\
\hline
$\lambda=2.5$&&&&&&&&\\
\hline
BTV(1/2,1/2) &&&&&&&&\\
$\mu$ & -0.225 & 0.011 & 0.363 & -0.207 & 0.037 & 0.382 & 0.930 & --\\
  $\sigma$ & 0.787 & 0.975 & 1.142 & 0.804 & 0.975 & 1.142 & 0.930 & --\\
  $\lambda$ & 1.069 & 2.233 & 3.760 & 1.187 & 2.330 & 3.999 & 0.931 & 0.001\\
Jeffreys &&&&&&&&\\
$\mu$ & -0.228 & 0.014 & 0.334 & -0.211 & 0.034 & 0.392 & 0.924 & --\\
  $\sigma$ & 0.790 & 0.974 & 1.143 & 0.804 & 0.977 & 1.145 & 0.926 & --\\
  $\lambda$ & 1.042 & 2.253 & 3.745 & 1.188 & 2.349 & 3.936 & 0.927 & 0.001\\
BTV(1,1) &&&&&&&&\\
$\mu$ & -0.207 & 0.029 & 0.370 & -0.194 & 0.052 & 0.427 & 0.918 & --\\
  $\sigma$ & 0.777 & 0.966 & 1.128 & 0.790 & 0.968 & 1.128 & 0.918 & --\\
  $\lambda$ & 0.954 & 2.174 & 3.617 & 1.088 & 2.236 & 3.800 & 0.911 & 0.002\\
\hline
$\lambda=5$&&&&&&&&\\
\hline
BTV(1/2,1/2) &&&&&&&&\\
$\mu$ & -0.141 & 0.003 & 0.157 & -0.132 & 0.009 & 0.172 & 0.934 & --\\
  $\sigma$ & 0.859 & 0.989 & 1.125 & 0.859 & 0.992 & 1.128 & 0.940 & --\\
  $\lambda$ & 2.773 & 4.640 & 7.651 & 3.013 & 4.942 & 8.602 & 0.942 & 9$\times10^{-9}$\\
Jeffreys &&&&&&&&\\
$\mu$ & -0.141 & 0.004 & 0.169 & -0.138 & 0.010 & 0.174 & 0.940 & --\\
  $\sigma$ & 0.854 & 0.987 & 1.126 & 0.859 & 0.992 & 1.132 & 0.933 & --\\
  $\lambda$ & 2.811 & 4.615 & 7.605 & 3.023 & 4.957 & 8.638 & 0.945 & 8$\times10^{-9}$\\
BTV(1,1) &&&&&&&&\\
$\mu$ & -0.131 & 0.012 & 0.179 & -0.123 & 0.018 & 0.184 & 0.936 & --\\
  $\sigma$ & 0.851 & 0.983 & 1.114 & 0.851 & 0.985 & 1.123 & 0.932 & --\\
  $\lambda$ & 2.704 & 4.429 & 7.301 & 2.883 & 4.754 & 8.134 & 0.939 & 8$\times10^{-9}$\\
\hline
\end{tabular}
\caption{Skew-logistic data for noninformative priors: $\mu=0,\sigma=1, n=200$.}
\label{table:SLO200}
\end{center}
\end{table}

\begin{table}[!htbp]
\begin{center}
\scriptsize
\begin{tabular}[h]{|c| c c c| c c c| c| c|}
\hline
Prior &  \multicolumn{3}{c}{MAP} & \multicolumn{3}{c}{Median} & Coverage & BF \\
\hline
 & 5\% & 50\% & 95\% & 5\% & 50\% & 95\% & &\\
  \hline
$\lambda=0$&&&&&&&&\\
\hline
BTV(1/2,1/2) &&&&&&&&\\
$\mu$ & -0.342 & -0.007 & 0.347 & -0.341 & 0.006 & 0.356 & 0.936 &--\\
  $\sigma$ & 0.850 & 1.021 & 1.197 & 0.868 & 1.038 & 1.213 & 0.950 &--\\
  $\lambda$ & -0.240 & 0.001 & 0.226 & -0.335 & 0.001 & 0.304 & 0.934 & 6.549\\
Jeffreys &&&&&&&&\\
$\mu$ & -0.343 & -0.007 & 0.350 & -0.347 & 0.006 & 0.349 & 0.939&-- \\
  $\sigma$ & 0.851 & 1.021 & 1.199 & 0.869 & 1.038 & 1.210 & 0.953 &--\\
  $\lambda$ & -0.244 & 0.001 & 0.221 & -0.330 & 0.004 & 0.303 & 0.936 & 6.333\\
BTV(1,1) &&&&&&&&\\
$\mu$ & -0.354 & -0.003 & 0.328 & -0.338 & 0.003 & 0.333 & 0.944 &--\\
  $\sigma$ & 0.853 & 1.021 & 1.200 & 0.868 & 1.035 & 1.212 & 0.952 &--\\
  $\lambda$ & -0.243 & 0.001 & 0.214 & -0.319 & 0.001 & 0.292 & 0.934 & 4.277\\
\hline
$\lambda=2.5$&&&&&&&&\\
\hline
BTV(1/2,1/2) &&&&&&&&\\
$\mu$ & -0.173 & 0.002 & 0.237 & -0.157 & 0.014 & 0.240 & 0.945 &--\\
  $\sigma$ & 0.781 & 0.969 & 1.195 & 0.799 & 0.983 & 1.216 & 0.952 &--\\
  $\lambda$ & 0.834 & 2.039 & 4.492 & 1.068 & 2.377 & 5.605 & 0.944 & 0.011\\
Jeffreys &&&&&&&&\\
$\mu$ & -0.173 & 0.001 & 0.249 & -0.155 & 0.012 & 0.238 & 0.938 &--\\
  $\sigma$ & 0.771 & 0.971 & 1.203 & 0.788 & 0.985 & 1.217 & 0.953 &--\\
  $\lambda$ & 0.841 & 2.065 & 4.552 & 1.060 & 2.386 & 5.598 & 0.940 & 0.010\\
BTV(1,1) &&&&&&&&\\
$\mu$ & -0.163 & 0.012 & 0.257 & -0.145 & 0.026 & 0.251 & 0.938 &--\\
  $\sigma$ & 0.772 & 0.958 & 1.192 & 0.788 & 0.974 & 1.200 & 0.944 &--\\
  $\lambda$ & 0.778 & 1.951 & 4.391 & 0.991 & 2.241 & 5.094 & 0.946 & 0.008\\
\hline
$\lambda=5$&&&&&&&&\\
\hline
BTV(1/2,1/2) &&&&&&&&\\
$\mu$ & -0.120 & 0.001 & 0.147 & -0.113 & 0.008 & 0.157 & 0.938 &--\\
  $\sigma$ & 0.786 & 0.974 & 1.187 & 0.801 & 0.983 & 1.194 & 0.947 &--\\
  $\lambda$ & 1.943 & 4.172 & 9.410 & 2.427 & 4.877 & 12.860 & 0.936 & 0.002\\
Jeffreys &&&&&&&&\\
$\mu$ & -0.128 & -0.000 & 0.153 & -0.117 & 0.009 & 0.159 & 0.943 &--\\
  $\sigma$ & 0.787 & 0.973 & 1.182 & 0.798 & 0.985 & 1.194 & 0.945 &--\\
  $\lambda$ & 1.950 & 4.128 & 10.446 & 2.381 & 4.868 & 12.906 & 0.936 & 0.003\\
BTV(1,1) &&&&&&&&\\
$\mu$ & -0.109 & 0.009 & 0.164 & -0.099 & 0.017 & 0.177 & 0.947 &--\\
  $\sigma$ & 0.772 & 0.962 & 1.159 & 0.787 & 0.974 & 1.182 & 0.947 &--\\
  $\lambda$ & 1.852 & 3.848 & 8.546 & 2.224 & 4.479 & 10.951 & 0.942 & 0.002\\
\hline
\end{tabular}
\caption{Skew-Laplace data for noninformative priors: $\mu=0,\sigma=1, n=100$.}
\label{table:SL100}
\end{center}
\end{table}

\begin{table}[!htbp]
\begin{center}
\scriptsize
\begin{tabular}[h]{|c| c c c| c c c| c| c|}
\hline
Prior &  \multicolumn{3}{c}{MAP} & \multicolumn{3}{c}{Median} & Coverage & BF \\
\hline
 & 5\% & 50\% & 95\% & 5\% & 50\% & 95\% & &\\
  \hline
$\lambda=0$&&&&&&&&\\
\hline
BTV(1/2,1/2) &&&&&&&&\\
$\mu$ & -0.204 & -0.002 & 0.225 & -0.209 & -0.002 & 0.223 & 0.941 & --\\
  $\sigma$ & 0.889 & 1.008 & 1.136 & 0.898 & 1.013 & 1.142 & 0.942 & --\\
  $\lambda$ & -0.147 & -0.000 & 0.149 & -0.183 & 0.000 & 0.185 & 0.947 & 10.585\\
Jeffreys &&&&&&&&\\
$\mu$ & -0.206 & -0.001 & 0.224 & -0.211 & -0.000 & 0.214 & 0.939  & --\\
  $\sigma$ & 0.884 & 1.007 & 1.133 & 0.895 & 1.014 & 1.141 & 0.939  & --\\
  $\lambda$ & -0.148 & 0.001 & 0.154 & -0.180 & 0.001 & 0.190 & 0.945  &  10.440\\
BTV(1,1) &&&&&&&&\\
$\mu$ & -0.208 & -0.002 & 0.211 & -0.206 & 0.001 & 0.210 & 0.941 & --\\
  $\sigma$ & 0.887 & 1.007 & 1.132 & 0.898 & 1.014 & 1.140 & 0.942 & --\\
  $\lambda$ & -0.138 & 0.000 & 0.148 & -0.180 & -0.000 & 0.189 & 0.945 & 6.852\\
\hline
$\lambda=2.5$&&&&&&&&\\
\hline
BTV(1/2,1/2) &&&&&&&&\\
$\mu$ & -0.126 & -0.004 & 0.155 & -0.119 & 0.003 & 0.155 & 0.945 & --\\
  $\sigma$ & 0.840 & 0.987 & 1.147 & 0.845 & 0.994 & 1.159 & 0.936 & --\\
  $\lambda$ & 1.307 & 2.311 & 3.849 & 1.429 & 2.469 & 4.217 & 0.941 & 7$\times10^{-6}$\\
Jeffreys &&&&&&&&\\
$\mu$ & -0.127 & -0.001 & 0.157 & -0.118 & 0.004 & 0.161 & 0.940 & --\\
  $\sigma$ & 0.836 & 0.982 & 1.148 & 0.851 & 0.992 & 1.154 & 0.939 & --\\
  $\lambda$ & 1.289 & 2.291 & 3.830 & 1.403 & 2.460 & 4.176 & 0.944 & 6$\times10^{-6}$\\
BTV(1,1) &&&&&&&&\\
$\mu$ & -0.119 & 0.004 & 0.159 & -0.110 & 0.009 & 0.160 & 0.945  & --\\
  $\sigma$ & 0.828 & 0.978 & 1.140 & 0.843 & 0.988 & 1.145 & 0.938  & --\\
  $\lambda$ & 1.263 & 2.210 & 3.738 & 1.382 & 2.385 & 4.045 & 0.945  & 5$\times10^{-6}$\\
\hline
$\lambda=5$&&&&&&&&\\
\hline
BTV(1/2,1/2) &&&&&&&&\\
$\mu$ & -0.084 & -0.001 & 0.097 & -0.080 & 0.002 & 0.096 & 0.951 & --\\
  $\sigma$ & 0.849 & 0.984 & 1.137 & 0.855 & 0.991 & 1.142 & 0.939 & --\\
  $\lambda$ & 2.817 & 4.596 & 7.956 & 3.036 & 4.900 & 8.621 & 0.952 & 7$\times10^{-8}$\\
Jeffreys &&&&&&&&\\
$\mu$ & -0.085 & -0.002 & 0.095 & -0.079 & 0.002 & 0.097 & 0.951 & --\\
  $\sigma$ & 0.847 & 0.987 & 1.142 & 0.853 & 0.990 & 1.142 & 0.940 & --\\
  $\lambda$ & 2.784 & 4.567 & 7.767 & 3.033 & 4.879 & 8.693 & 0.945 & 1$\times10^{-7}$\\
BTV(1,1) &&&&&&&&\\
$\mu$ & -0.080 & 0.004 & 0.099 & -0.074 & 0.006 & 0.102 & 0.951 & --\\
  $\sigma$ & 0.840 & 0.978 & 1.128 & 0.849 & 0.986 & 1.135 & 0.932 & --\\
  $\lambda$ & 2.615 & 4.427 & 7.547 & 2.859 & 4.744 & 8.132 & 0.943 & 7$\times10^{-8}$\\
\hline
\end{tabular}
\caption{Skew-Laplace data for noninformative priors: $\mu=0,\sigma=1, n=200$.}
\label{table:SL200}
\end{center}
\end{table}

\newpage
 \setlength{\bibsep}{ 2.5pt}
\bibliographystyle{plainnat}
\bibliography{references}

\end{document}